\documentstyle[epsfig,aps,pre]{revtex}
\tightenlines
\def\beq{\begin{equation}}
\def\beqa{\begin{eqnarray}}
\def\eeq{\end{equation}}
\def\eeqa{\end{eqnarray}}

\def\one{{(1)}}
\def\two{{(2)}}
\def\three{{(3)}}

\begin{document}
\title{Virial coefficients and equations of state for mixtures of hard discs, 
hard
spheres and hard hyperspheres }
\author{A. Santos\cite{andres}}
\address{Department of Physics, University of Florida,\\
Gainesville, FL 32611, U.S.A.}
\author{S. B. Yuste\cite{santos}}
\address{Institute for Nonlinear Science and Department of Chemistry, University 
of California,\\
San Diego, La Jolla, CA 92093-0340, U.S.A.}
\author{M. L\'{o}pez de Haro\cite{mariano}}\address{Centro de Investigaci\'on en 
Energ\'{\i}a, UNAM,\\
Temixco, Morelos 62580, M\'{e}xico}

\date{\today}

\maketitle

\begin{abstract}
The composition-independent virial coefficients of a $d$-dimensional binary
mixture of { (additive)} hard  hyperspheres following from a recent proposal 
for
the equation of state of the mixture [{\sc Santos, A., Yuste, S. B., and 
L\'{o}pez de Haro, M.}, 1999, {\em Molec. Phys.}, {\bf 96}, 1] are examined. 
Good
agreement between theoretical estimates and available exact or numerical
results is
found for $d=2$, 3, 4 and $5$, { except for mixtures whose components are very disparate in size.}
A slight modification that remedies this
deficiency is introduced  { and the resummation of the associated virial series is carried out,
leading} to a new proposal for the equation of
state. The case of binary hard-sphere mixtures ($d=3$) is analyzed in some 
detail.
\end{abstract}

\section{Introduction}

\label{sec1} Depending on the nature of the independent variables, it is
well known that the full thermodynamic description of a given system
requires the availability of either one of the thermodynamic potentials or
of the corresponding equations of state. In the case of classical fluids,
the usual expression for the equation of state (EOS) is a relationship among
the pressure $p$, the density $\rho $ and the temperature $T$ of the fluid.
However, except for very few ideal systems, the explicit (exact) form of
this relationship is not known in general. Therefore, given the fact that an
experimental determination of the EOS for every particular fluid is of
course not practical, researchers have concentrated their efforts in
proposing \ ``reasonable'' empirical or semi-empirical approximations.
Perhaps the neatest example of such an approach, already produced over a
century ago and providing an essentially correct qualitative picture of the
thermodynamic properties of gases and liquids, is the celebrated van der
Waals EOS\cite{vdW1873}. An alternative proposal, first introduced by
Kammerlingh Onnes\cite{KO} { as a
mathematical representation of experimental results on the equation of state
of gases and liquids,} is the so called virial equation of state. Written for
the compressibility factor $Z\equiv {p}/\rho k_{B}T$ (where $k_{B}$ is the
Boltzmann constant) as a power series in $\rho $, it provides an expansion
that gives the deviation from ideal gas behaviour in ascending powers of the
density, namely

\begin{equation}
Z=1+\sum\limits_{n=2}^{\infty }B_{n}\rho ^{n-1}.  \label{1v}
\end{equation}
Here, the $B_{n}$ are the virial coefficients. For a simple fluid, the
virial coefficients are in general functions of $T$ alone, while in the case
of mixtures they also depend on composition\cite{MS69}. An interesting
aspect of the virial equation of state, discovered many years after it was
introduced, is the fact that it can be derived rigorously using statistical
mechanics\cite{MM40}. This in turn implies that the virial coefficients are
not merely empirical constants but are rather related to intermolecular
interactions in a well-defined manner. It is unfortunate, however, that in
general the actual computation of the virial coefficients is a formidable
task and that the radius of convergence of the series in Eq. (\ref{1v}) is
not known.

In the case of hard-core fluids (rods, discs, spheres and hyperspheres) the
virial coefficients are easier to compute than those corresponding to any
other intermolecular interactions. Even in these model systems and with the
exception of hard rods, where the exact EOS and corresponding virial
coefficients are known, only the first few of such coefficients are
available so far\cite{vvc1,JvR93,LB82,J82} because the number of cluster
integrals involved increases very rapidly with the order of the coefficient.
The availability is scarcer for mixtures than for simple fluids, although
recently there has been renewed interest in partially remedying this
deficiency 
\cite{KM75,BXHB88,SFG96,EACG97,SFG97,EAGB98,W98a,WSG98,B98,W99a,EAGB00}. The 
limited knowledge of the virial
coefficients has proved useful, however, in the sense that proposals for the
EOS of these systems may be judged amongst other things by how well the
virial coefficients arising in such proposals compare with the exact values.
The rationale here is that a ``good'' theoretical EOS should lead to an
accurate prediction of the value of the virial coefficients. Alternatively,
the few available virial coefficients may also be profitably used to
construct a rational (Pad\'{e} or Levin)\cite{JvR93,EW84,EL85,S94,CB98,MV99}
aproximation to the EOS or to guide in the construction of a theoretical
EOS, as was the case with the Carnahan-Starling (CS) EOS\cite{CS69} for a { 
simple} hard-sphere fluid.

A variety of (approximate) EOS for { simple fluids composed of
hard discs and hard spheres are available in the literature\cite{EOSDE} and
some work has also been reported on simple fluids of hard hyperspheres 
\cite{BC87,SM90,LM90,BMC99,S00}. Very} recently we introduced a simple recipe 
to derive
the compressibility factor of a multicomponent mixture of $d$-dimensional
hard hyperspheres in terms of that of the corresponding single component
system\cite{SYH99}. A straightforward consequence of such a recipe is that
one can easily derive the explicit expressions for the virial coefficients
of the mixture in terms of those of the single component fluid for all $d$.
The major aim of this paper is to further assess the usefulness of our
recipe by comparing the predictions of the values of the virial coefficients
of binary mixtures of hard discs, hard spheres and hard hyperspheres in $d=4$
and $d=5$ with those obtained through Monte Carlo integration. Along the way
we will introduce slight modifications to our original proposal and consider
an alternative EOS to cope with mixtures very disparate in size. This new
EOS is consistent with the forms suggested by Wheatley for the virial
coefficients of hard discs \cite{W98b} and hard spheres \cite{W99b}.

The paper is organized as follows. In the next Section and in order to make
the paper self-contained, we will briefly recall the main results of our
previous work and write down explicitly the composition-independent virial
coefficients of a $d$-dimensional binary mixture of { (additive)} hard 
hyperspheres. Section \ref{sec3} deals with a new proposal for the EOS of
the binary mixture that attempts to incorporate additional features not
present in the original recipe. In Section \ref{sec4} we focus on the case
of hard spheres because of its intrinsic interest. We close the paper in
Section \ref{sec5} with an assessment of the different results. Some
mathematical details of our derivations are presented in two Appendices.

\section{Approximate equation of state  { and virial coefficients} 
for a binary mixture of 
\lowercase{$d$}-dimensional hard spheres}

\label{sec2} In this Section we first consider a multicomponent mixture of 
$d$-dimensional (additive) hard hyperspheres and then restrict ourselves to
{\em binary\/} mixtures of hard hyperspheres in $d$ dimensions.

Let the number of components be $N$, the total number density of the mixture
be $\rho $, the set of mole fractions be $\{x_{i}\}$, and the set of
diameters be $\{\sigma _{i}\}$, $(i=1,2,\ldots,N)$. The packing fraction is 
$\eta
=\sum_{i=1}^{N}\eta _{i}=v_{d}\rho \langle \sigma ^{d}\rangle $, where $\eta
_{i}=$ $v_{d}\rho _{i}\sigma _{i}^{d}$ is the partial packing fraction due
to species $i$, $\rho _{i}=\rho x_{i}$ is the partial number density
corresponding to species $i$, $v_{d}=(\pi /4)^{d/2}/\Gamma (1+d/2)$ is the
volume of a $d$-dimensional sphere of unit diameter and $\langle \sigma
^{n}\rangle \equiv \sum_{i=1}^{N}x_{i}\sigma _{i}^{n}$. In previous work 
\cite{SYH99} we proposed a simple equation of state for the mixture, 
$Z_{\text{m}}(\eta )$, { consistent with a given EOS for a single component
system, $Z_{\text{s}}(\eta )$, at the same packing fraction $\eta $,  namely} 
\begin{equation}
Z_{\text{m}}(\eta )=1+\left[ Z_{\text{s}}(\eta )-1\right] 2^{1-d}\Delta _{0}+
\frac{\eta }{1-\eta }\left( 1-\Delta _{0}+\frac{1}{2}\Delta _{1}\right) ,
\label{4.1}
\end{equation}
with 
\begin{equation}
\Delta _{p}\equiv \frac{\langle \sigma ^{d+p-1}\rangle }{\langle \sigma
^{d}\rangle ^{2}}\sum_{m=p}^{d-1}\frac{(d+p-1)!}{m!(d+p-1-m)!}\langle \sigma
^{m-p+1}\rangle \langle \sigma ^{d-m}\rangle .  \label{2}
\end{equation}
In the one-dimensional case, Eq.\ (\ref{4.1}) yields the {\em exact\/}
result $Z_{\text{m}}(\eta )=Z_{\text{s}}(\eta )$. Further, for binary
mixtures in $d=2$, $3$, $4$ and $5$, it proved to be very satisfactory when
a reasonably accurate $Z_{\text{s}}(\eta )$ was taken \cite
{MV99,SYH99,CCHW00,AGH}.

{}{ From the virial expansion $Z_{\text{s}}(\eta )=
1+\sum_{n=2}^{\infty }b_{n}\eta ^{n-1}$, where $b_{n}$ are (reduced) virial
coefficients, and according} to Eqs.\ (\ref{1v})--(\ref{2}), the $n$-th virial coefficient of
the mixture is given by 
\begin{equation}
B_{n}=v_{d}^{n-1}\langle \sigma ^{d}\rangle ^{n-1}\left( 2^{1-d}\Delta
_{0}b_{n}+1-\Delta _{0}+\frac{1}{2}\Delta _{1}\right) ,  \label{1}
\end{equation}
and explicitly for $n\geq 2$ 
\begin{equation}
B_{n}=v_{d}^{n-1}\langle \sigma ^{d}\rangle ^{n-3}\sum_{m=0}^{d}\frac{(d-1)!
}{m!(d-m)!}\langle \sigma ^{d-m}\rangle \left[ (2^{1-d}b_{n}-1)(d-m)\langle
\sigma ^{m+1}\rangle \langle \sigma ^{d-1}\rangle +\frac{d}{2}\langle \sigma
^{m}\rangle \langle \sigma ^{d}\rangle \right] .  \label{3}
\end{equation}

Now we particularize to {\em binary\/} mixtures. In that case, it is useful
to define composition-independent coefficients $B_{n_{1},n_{2}}$ as 
\begin{equation}
B_{n}=\sum_{n_{1}=0}^{n}B_{n_{1},n-n_{1}}\frac{n!}{n_{1}!(n-n_{1})!}
x_{1}^{n_{1}}x_{2}^{n-n_{1}}.  \label{4}
\end{equation}
Our objective is to use our model to get approximate expressions of $
B_{n_{1},n_{2}}$ as explicit functions of $\sigma _{1}$, $\sigma _{2}$, and $
d$. In Appendix \ref{appA} it is shown that the result is 
\begin{eqnarray}
B_{n_{1},n_{2}} &=&{v_{d}^{n-1}\sigma _{1}^{d(n-1)}\alpha ^{d(n_{2}-1)}}
\left[ C_{n_{1},n_{2}}^{{(1)}}\alpha ^{d}+C_{n_{1},n_{2}}^{{(2)}}\alpha
^{d-1}+C_{n_{1},n_{2}}^{{(3)}}(1+\alpha )^{d-1}\right.  \nonumber \\
&&\left. +C_{n_{2},n_{1}}^{{(3)}}\alpha (1+\alpha )^{d-1}+C_{n_{2},n_{1}}^{{
(2)}}\alpha +C_{n_{2},n_{1}}^{{(1)}}\right] ,  \label{11}
\end{eqnarray}
where $n\equiv n_{1}+n_{2}$, $\alpha\equiv \sigma_2/\sigma_1$ and 
\begin{equation}
C_{n_{1},n_{2}}^{{(1)}}=\frac{n_{1}(n_{1}-1)\left[
(n_{1}-2)b_{n}+2^{d-1}n_{2}\right] }{n(n-1)(n-2)},  \label{11.1}
\end{equation}
\begin{equation}
C_{n_{1},n_{2}}^{{(2)}}=\frac{n_{1}n_{2}(n_{1}-1)\left( b_{n}-2^{d-1}\right) 
}{n(n-1)(n-2)},  \label{11.2}
\end{equation}
\begin{equation}
C_{n_{1},n_{2}}^{{(3)}}=\frac{n_{1}n_{2}\left[
(n_{2}-1)2^{2-d}b_{n}+n_{1}-n_{2}\right] }{n(n-1)(n-2)}.  \label{11.3}
\end{equation}
The special case of $n=2$ is obtained by {\em first\/} setting $
b_{n}=2^{d-1} $ and {\em then\/} setting $n=2$. The result is 
\begin{equation}
C_{n_{1},n_{2}}^{{(1)}}=2^{d-2}n_{1}(n_{1}-1), \quad C_{n_{1},n_{2}}^{{(2)}
}=0, \quad C_{n_{1},n_{2}}^{{(3)}}=\frac{1}{2}n_{1}n_{2}.  \label{11.4}
\end{equation}
This gives the {\em exact\/} second virial coefficient.

Equation (\ref{11}) expresses the reduced coefficient 
\begin{equation}
B_{n_{1},n_{2}}^{\ast }(\alpha )\equiv B_{n_{1},n_{2}}(\sigma _{1},\sigma
_{2})\sigma _{1}^{-d(n-1)}\alpha ^{-d(n_{2}-1)}  \label{11.5}
\end{equation}
as a polynomial in $\alpha $ of degree $d$. Although it is known that the 
{\em exact\/} coefficients do not have in general such a polynomial
structure \cite{W99b}, most of the proposals to date for $d=2$ and $d=3$ are
polynomials \cite{W98a,WSG98,W99a,CB98,MV99,W98b,W99b}. Let us see which
consistency conditions the approximation (\ref{11})--(\ref{11.3}) fulfills.
First, it verifies the obvious symmetry property $B_{n_{1},n_{2}}(\sigma
_{1},\sigma _{2})=B_{n_{2},n_{1}}(\sigma _{2},\sigma _{1})$, which implies
that $B_{n_{1},n_{2}}^{\ast }(\alpha )=\alpha ^{d}B_{n_{2},n_{1}}^{\ast
}(1/\alpha )$. In addition, the coefficients (\ref{11.1})--(\ref{11.3})
satisfy the properties 
\begin{equation}
C_{n_{1},n_{2}}^{{(1)}}+C_{n_{2},n_{1}}^{{(2)}}+2^{d-1}C_{n_{2},n_{1}}^{{(3)}
}=\frac{n_{1}}{n}b_{n},  \label{18}
\end{equation}
\begin{equation}
dC_{n_{1},n_{2}}^{{(1)}}+(d-2)C_{n_{1},n_{2}}^{{(2)}
}+2^{d-1}C_{n_{2},n_{1}}^{{(3)}}=\frac{n_{1}}{n(n-1)}\left[
2^{d-1}n_{2}+d(n_{1}-1)b_{n}\right] ,  \label{26}
\end{equation}
\begin{equation}
C_{n,0}^{{(1)}}=b_{n},\quad C_{0,n}^{{(1)}}=C_{n,0}^{{(2)}}=C_{0,n}^{{(2)}
}=C_{n,0}^{{(3)}}=C_{0,n}^{{(3)}}=0.  \label{17}
\end{equation}
Equation (\ref{18}) implies that 
\begin{equation}
C_{n_{1},n_{2}}^{{(1)}}+C_{n_{2},n_{1}}^{{(1)}}+C_{n_{1},n_{2}}^{{(2)}
}+C_{n_{2},n_{1}}^{{(2)}}+2^{d-1}\left[ C_{n_{1},n_{2}}^{{(3)}
}+C_{n_{2},n_{1}}^{{(3)}}\right] =b_{n},  \label{23}
\end{equation}
which guarantees that 
\begin{equation}
B_{n_{1},n_{2}}^{\ast }(\alpha =1)=v_{d}^{n-1}b_{n},  \label{41}
\end{equation}
i.e., if both species have the same size we recover the one-component case.
The same situation appears if $x_{1}=0$ or $x_{2}=0$, which means that $
B_{n,0}^{\ast }(\alpha )=v_{d}^{n-1}b_{n}\alpha ^{d}$ and $B_{0,n}^{\ast
}(\alpha )=v_{d}^{n-1}b_{n}$. This is verified as a consequence of Eq.\ (\ref
{17}). A subtler consistency condition is \cite{W99b} 
\begin{equation}
\left. \frac{\partial B_{n_{1},n_{2}}^{\ast }}{\partial \alpha }\right|
_{\alpha =1}=d\frac{n_{1}}{n}v_{d}^{n-1}b_{n},  \label{21}
\end{equation}
which ensures that the derivative of the excess free energy of mixing with
respect to $\sigma _{2}$ is zero when the spheres are of the same size \cite
{W99b}. In the approximation (\ref{11}), Eq.\ (\ref{21}) requires 
\begin{equation}
dC_{n_{1},n_{2}}^{{(1)}}+(d-1)C_{n_{1},n_{2}}^{{(2)}
}+2^{d-2}(d-1)C_{n_{1},n_{2}}^{{(3)}}+2^{d-2}(d+1)C_{n_{2},n_{1}}^{{(3)}
}+C_{n_{2},n_{1}}^{{(2)}}=d\frac{n_{1}}{n}b_{n}.  \label{22}
\end{equation}
If we interchange the roles of $n_{1}$ and $n_{2}$ in Eq.\ (\ref{22}) and
add the original and the transformed equations, we get Eq.\ (\ref{23}). If,
instead, we subtract both equations, we obtain 
\begin{equation}
d\left[ C_{n_{1},n_{2}}^{{(1)}}-C_{n_{2},n_{1}}^{{(1)}}\right] +(d-2)\left[
C_{n_{1},n_{2}}^{{(2)}}-C_{n_{2},n_{1}}^{{(2)}}\right] -2^{d-1}\left[
C_{n_{1},n_{2}}^{{(3)}}-C_{n_{2},n_{1}}^{{(3)}}\right] =d\frac{n_{1}-n_{2}}{n
}b_{n}.  \label{25}
\end{equation}
Thus, enforcement of Eq.\ (\ref{22}) is equivalent to enforcement of Eq.\ 
(\ref{25}). The property (\ref{26}) guarantees that the condition (\ref{25})
is fulfilled by the coefficients (\ref{11.1})--(\ref{11.3}).

The most stringent conditions appear in the limits $\alpha \rightarrow 0$
and $\alpha \rightarrow \infty $. In that case, Eq.\ (\ref{11}) yields 
\beq
B_{n_{1},n_{2}}^{\ast }(\alpha ) =
\left\{
\begin{array}{ll}
v_{d}^{n-1}\left[ C_{n_{1},n_{2}}^{{(3)}
}+C_{n_{2},n_{1}}^{{(1)}}\right]& (\alpha \rightarrow 0),  \\
v_{d}^{n-1}\alpha ^{d}\left[ 
C_{n_{1},n_{2}}^{{(1)}}+C_{n_{2},n_{1}}^{{(3)}}\right]& (\alpha \rightarrow 
\infty ). 
\end{array}\right.
 \label{12}
\eeq
On the other hand, the {\em exact\/} result is \cite{W98b,W99b,S99} 
\beq
B_{n_{1},n_{2}}^{\ast }(\alpha ) =
\left\{
\begin{array}{ll}v_{d}^{n-1}\frac{n_{2}}{n}b_{n_{2}}
& (\alpha \rightarrow 0),  \\
v_{d}^{n-1}\alpha ^{d}\frac{n_{1}}{n}b_{n_{1}}
& (\alpha \rightarrow 
\infty ). 
\end{array}\right.
 \label{13}
\eeq
The above condition means that in the limit of infinite size asymmetry, the
smaller spheres contribute to the total pressure as if they were a
one-component system in a free volume equal to the total volume minus the
volume occupied by the larger spheres. Our approximation (\ref{11})--(\ref
{11.3}) gives the correct forms for the asymptotic behaviours of $
B_{n_{1},n_{2}}^{\ast }$, but not the correct coefficients (except for $
n_{2}=1$ in the limit $\alpha \rightarrow 0$ and for $n_{1}=1$ in the limit $
\alpha \rightarrow \infty $, apart from the trivial cases $n_{1}=0,n$ and $
n_{2}=0,n$). In fact, in our approximation the virial coefficient $B_{n}$
depends parametrically on $b_{n}$ only, and not on the previous coefficients 
$b_{n_{1}},n_{1}\leq n$, as is needed in Eq.\ (\ref{13}). This is a
consequence of the form of our approximation  in which the
compressibility factor of the mixture at a given packing fraction $\eta $ is
expressed in terms of the compressibility factor of the monodisperse system
at the {\em same\/} packing fraction.

Another consistency condition is \cite{W99b} 
\begin{equation}
\left. \frac{\partial B_{n-1,1}^{\ast }}{\partial \alpha }\right| _{\alpha
=0}=v_{d}^{n-1}d\frac{n-1}{n}.  \label{13bis}
\end{equation}
This comes from the exact value of the derivative of the excess free energy
with respect to molecular size in the limit of infinite size asymmetry \cite
{THM99}. In general, this condition is not verified by our approximation
either. We consider the condition (\ref{13}) to be more important than (\ref
{13bis}) because it is not restricted to $n_{2}=1$ (or $n_{1}=1$) and also
because it refers to the leading term in the limit of infinite size
asymmetry, while the condition (\ref{13bis}) refers to the sub-leading term.

\begin{table}[tbp]
\caption{Reduced virial coefficients $b_{n}$ for the one-component case. }
\label{table1}
\begin{tabular}{ccccc}
$n$ & $d=2$\tablenote{Ref.\ \protect\cite{JvR93}} & $d=3^{\text{a}}$ & 
$d=4$\tablenote{Refs.\ \protect\cite{LB82}, \protect\cite{LM90} and
\protect\cite{BMC99}} & $d=5^{\text{b}}$ \\ 
\tableline 2 & 2 & 4 & 8 & 16 \\ 
3 & 3.12801775 & 10 & 32.4057594 & 106 \\ 
4 & 4.25785446 & 18.36477 & 77.7451797 & 311.18341 \\ 
5 & 5.336897 & 28.2245 & 145.9 & 843.4 \\ 
6 & 6.3626 & 39.739 & 252.0 & 988 \\ 
7 & 7.351 & 53.539 &  &  \\ 
8 & 8.338 & 70.78 &  & 
\end{tabular}
\end{table}
Table \ref{table1} gives the known values of the coefficients $b_{n}$ for $
2\leq d\leq 5$. {}From them, Eqs.\ (\ref{11})--(\ref{11.4}) can be used to
estimate the coefficients $B_{n_{1},n_{2}}^*$ for any value of $\alpha $
in these dimensions. We will come back to this point later on.
The values of Table \ref{table1} can also be used to check to which degree
Eq.\ (\ref{13}) is violated by the approximation (\ref{11})--(\ref{11.3}).
The less favorable case corresponds to $n_{1}=1$ in the limit $\alpha
\rightarrow 0$ (or, equivalently, $n_{2}=1$ in the limit $\alpha \rightarrow
\infty $). Let us then define 
\begin{equation}
R_{n}\equiv \frac{C_{1,n-1}^{{(3)}}+C_{n-1,1}^{{(1)}}}{[(n-1)/n]b_{n-1}}.
\label{14}
\end{equation}
According to Eqs.\ (\ref{11.1}) and (\ref{11.3}), we have 
\begin{equation}
R_{n}=\frac{\left( 2^{2-d}+n-3\right) b_{n}+2^{d-1}-1}{(n-1)b_{n-1}}.
\label{15}
\end{equation}
Table \ref{table2} shows the values of $R_{n}$ that can be obtained from the
known virial coefficients. In the cases $d=2$ and $d=3$ the known values of
the ratio $R_{n}$ differ from 1 less than 5\%. In fact, by assuming $
R_{n}\simeq 1$, Eq.\ (\ref{15}) can be used to estimate $b_{9}$ and $b_{10}$. 
This gives $b_{9}\simeq 9.39$ ($d=2$), $b_{9}\simeq 86.7$ ($d=3$), $
b_{10}\simeq 10.44$ ($d=2$) and $b_{10}\simeq 103.6$ ($d=3$). Estimates
based on Pad\'{e} approximants are \cite{JvR93} $b_{9}\simeq 9.37$ ($d=2$), $
b_{9}\simeq 93.1$ ($d=3$), $b_{10}\simeq 10.55$ ($d=2$) and $b_{10}\simeq
123.2$ ($d=3$). On the other hand, the deviations of $R_{n}$ from 1 become
more important as the dimensionality increases and are already relatively
large for $d=5$. Thus, at least for those high dimensionalities, it appears
to be convenient to modify the prescription (\ref{11}) as to satisfy the
requirement (\ref{13}). This is the subject of the next Section, where
we will also discuss in more detail how good the estimates of 
$B_{n_{1},n_{2}}^{\ast }$, as obtained after using Eqs.\ (\ref{11})--(\ref{11.4}) 
with the values of the coefficients $b_{n}$   given in
Table \ref{table1}, are  for any value of $\alpha $ and $2\leq d\leq 5$.  
\begin{table}[tbp]
\caption{Ratio $R_{n}$, as given by Eq. (\ref{15}).}
\label{table2}
\begin{tabular}{ccccc}
$n$ & $d=2$ & $d=3$ & $d=4$ & $d=5$ \\ 
\tableline 3 & 1.032 & 1 & 0.944 & 0.883 \\ 
4 & 1.014 & 1.018 & 1.072 & 1.148 \\ 
5 & 0.999 & 1.001 & 1.078 & 1.452 \\ 
6 & 0.991 & 1.007 & 1.132 & 0.736 \\ 
7 & 0.989 & 1.023 &  &  \\ 
8 & 0.992 & 1.047 &  & 
\end{tabular}
\end{table}

\section{A new proposal for the equation of state of a binary mixture}
\label{sec3} 
In order to account for the conditions imposed by Eq. (\ref{13}) 
without greatly sacrificing the simplicity of our original recipe, let us
now explore the possibility of keeping the structure (\ref{11}), but with
expressions for the coefficients different from (\ref{11.1})--(\ref{11.3}).
The requirement (\ref{13}) implies that 
\begin{equation}
C_{n_{1},n_{2}}^{{(3)}}+C_{n_{2},n_{1}}^{{(1)}}=\frac{n_{2}}{n}b_{n_{2}}.
\label{16}
\end{equation}
In addition, the condition (\ref{41}) yields Eq.\ (\ref{23}). Note that Eq.\
(\ref{18}) is more restrictive than Eq.\ (\ref{23}). Since we want our
modified expressions to remain as close as possible to the original ones, we
will take (\ref{18}) as a second condition. {}From Eqs.\ (\ref{16}) and (\ref
{18}) we have 
\begin{equation}
C_{n_{1},n_{2}}^{{(1)}}=\left( 2^{d-1}-1\right) ^{-1}\left[ \frac{n_{1}}{n}
\left( 2^{d-1}b_{n_{1}}-b_{n}\right) +C_{n_{2},n_{1}}^{{(2)}}\right] ,
\label{19}
\end{equation}
\begin{equation}
C_{n_{1},n_{2}}^{{(3)}}=\left( 2^{d-1}-1\right) ^{-1}\left[ \frac{n_{2}}{n}
\left( b_{n}-b_{n_{2}}\right) -C_{n_{1},n_{2}}^{{(2)}}\right] .  \label{20}
\end{equation}
We need a third condition to close the problem. Since $C_{n_{1},n_{2}}^{{(2)}
}$ does not enter into Eq.\ (\ref{12}) and in agreement with our philosophy
of departing from the original formulation as less as possible, one
possibility is to keep Eq.\ (\ref{11.2}); { however}, the resulting 
approximation is
consistent with the conditions (\ref{17}) but does not satisfy the
requirement (\ref{22}). If we enforce the fulfillment of Eq.\ (\ref{22})
[or, equivalently, of Eq.\ (\ref{25})], then $C_{n_{1},n_{2}}^{{(2)}}$ must
have the following form 
\begin{equation}
C_{n_{1},n_{2}}^{{(2)}}=D_{n_{1},n_{2}}-\frac{n_{1}}{n}\frac{b_{n}-b_{n_{1}}
}{2^{2-d}-1},  \label{24}
\end{equation}
where $D_{n_{1},n_{2}}$ must be {\em symmetric\/}, but otherwise remains so
far unknown. To close the problem, we impose the more restrictive property 
(\ref{26}). This gives rise to 
\begin{eqnarray}
D_{n_{1},n_{2}} &=&\frac{1}{n\left[ 2+2^{d-1}(d-3)\right] }\left[ \left(
2^{d-1}-1\right) \left( 2^{d-1}-db_{n}\right) \frac{n_{1}n_{2}}{n-1}\right. 
\nonumber \\
&&\left. +\frac{2^{d-1}-d}{1-2^{2-d}}\left(
nb_{n}-n_{1}b_{n_{1}}-n_{2}b_{n_{2}}\right) \right] .  \label{27}
\end{eqnarray}
Equations (\ref{19})--(\ref{27}) close the modified version of our
approximation.

It must be noted that Eqs.\ (\ref{24}) and (\ref{27}) are meaningless if $
d=2 $. This is related to the fact that if $d=2$ it is impossible to enforce
conditions (\ref{41}), (\ref{21}) and (\ref{13}) simultaneously on a
polynomial of second degree for $B_{n_{1},n_{2}}^{\ast }$, as the number of
conditions exceeds the number of unknowns. As a matter of fact, Eqs. (\ref
{19}) and ({\ref{20}) are enough to uniquely determine $B_{n_{1},n_{2}}^{
\ast }$ in the case $d=2$, irrespective of the value of $C_{n_{1},n_{2}}^{{
(2)}}$. The result is 
\begin{equation}
B_{n_{1},n_{2}}^{\ast }=v_{2}^{n-1}\left[ \frac{n_{2}}{n}b_{n_{2}}+\left(
b_{n}-\frac{n_{1}}{n}b_{n_{1}}-\frac{n_{2}}{n}b_{n_{2}}\right) \alpha +\frac{
n_{1}}{n}b_{n_{1}}\alpha ^{2}\right] .  \label{42}
\end{equation}
This was the form proposed by Wheatley \cite{W98b} by imposing conditions 
(\ref{41}) and (\ref{13}). However, condition (\ref{21}) is not verified
unless $n_{1}(b_{n}-b_{n_{1}})=n_{2}(b_{n}-b_{n_{2}})$, which is only true
if $b_{n}=n$, i.e., if one considers the scaled particle theory \cite{SPT}}.
For the sake of comparison, the form that stems from our original
approximation, Eqs.\ (\ref{11})--(\ref{11.3}), is (for $d=2$) 
\begin{equation}
B_{n_{1},n_{2}}^{\ast }=v_{2}^{n-1}\frac{n_{1}n_{2}}{n(n-1)}\left[ \frac{
n_{2}-1}{n_{1}}b_{n}+1+2(b_{n}-1)\alpha +\left( \frac{n_{1}-1}{n_{2}}
b_{n}+1\right) \alpha ^{2}\right] .  \label{47}
\end{equation}

In summary, the modified expressions for the virial coefficients are given
by Eq.\ (\ref{42}) for $d=2$ and by Eqs.\ (\ref{11}) and (\ref{19})--(\ref
{27}) for $d\geq 3$. We are now in a position to compare the theoretical
predictions we have provided in this and the previous Section with the
available Monte Carlo data for the virial coefficients. We note that the
coefficients $B_{n_{1},n_{2}}$ corresponding to $n\leq 5$ have been
recently evaluated for $d=2$\cite{SFG96,W98a}, $d=3$ 
\cite{EACG97,SFG97,EAGB98,WSG98}, $d=4$\cite{EAGB00} and $d=5$ \cite{EAGB00}.
Further, the coefficients corresponding to $n=6$ for $d=2$ are also
known \cite{W99a}. { Because of its special physical interest and} for 
reasons that will become apparent later, the case of
hard spheres ($d=3$) will be addressed separately in Section \ref{sec4}.

We begin our assessment of our estimates of $B_{n_{1},n_{2}}$ with the
binary mixture of hard discs. This implies using Eq.\ (\ref{47}) for the
original recipe or using Eq.\ (\ref{42}) in the modified version, which
coincides with Wheatley's proposal \cite{W98b}. In Figs.\ {1}--\ref{fig4} 
we compare the results of taking these two routes with the ones of
Refs.\ \cite{BXHB88,SFG96,W98a,W99a}. As clearly seen in the figures, the
theoretical prescriptions do a very good job for the whole range of $\alpha $, 
the modified version (\ref{42}) being almost perfect.

\begin{figure}[tbp]
\begin{center}
\parbox{0.8\textwidth}{
\epsfxsize=\hsize \epsfbox{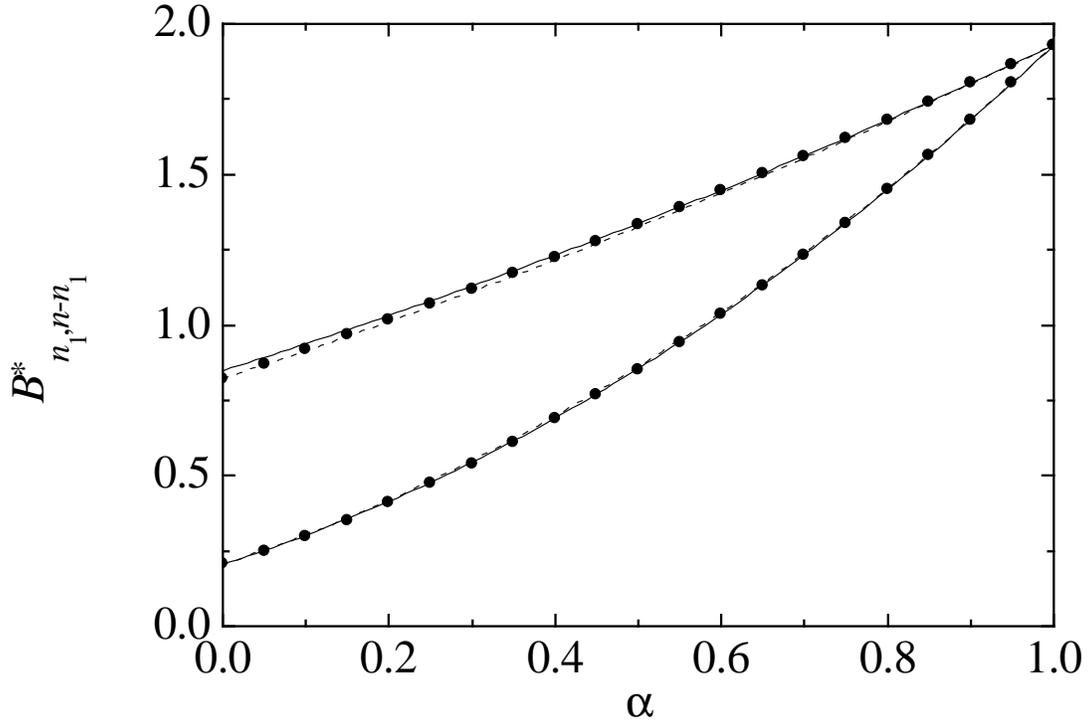}}
\caption{Reduced composition-independent virial coefficients $
B_{n_1,n-n_1}^* $ as functions of the size ratio $\protect\alpha$ for $d=2$, 
$n=3$, and, from top to bottom, $n_1=1$ and 2. The circles are exact results 
\protect\cite{BXHB88}, the solid line is the theoretical prediction (\ref{47}) 
and the dashed line is the theoretical prediction (\ref{42}). }
\end{center}
\label{fig1}
\end{figure}

\begin{figure}[tbp]
\begin{center}
\parbox{0.8\textwidth}{
\epsfxsize=\hsize \epsfbox{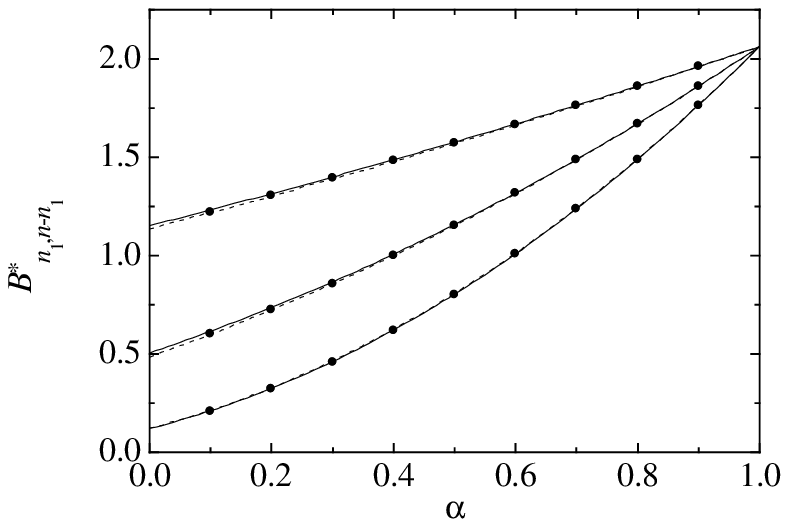}}
\caption{Reduced composition-independent virial coefficients $
B_{n_1,n-n_1}^* $ as functions of the size ratio $\protect\alpha$ for $d=2$, 
$n=4$, and, from top to bottom, $n_1=1$, 2 and 3. The circles are ``exact''
Monte Carlo results \protect\cite{SFG96}, the solid line is the theoretical
prediction (\ref{47}) and the dashed line is the theoretical prediction (\ref
{42}). }
\end{center}
\label{fig2}
\end{figure}

\begin{figure}[tbp]
\begin{center}
\parbox{.8\textwidth}{
\epsfxsize=\hsize \epsfbox{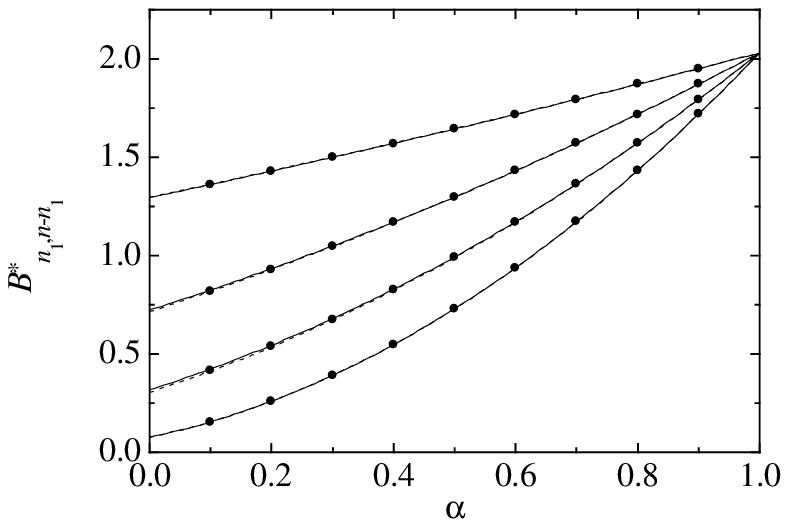}}
\caption{Reduced composition-independent virial coefficients $
B_{n_1,n-n_1}^* $ as functions of the size ratio $\protect\alpha$ for $d=2$, 
$n=5$, and, from top to bottom, $n_1=1$, 2, 3 and 4. The circles are
``exact'' Monte Carlo results \protect\cite{W98a}, the solid line is the
theoretical prediction (\ref{47}) and the dashed line is the theoretical
prediction (\ref{42}). }
\label{fig3}
\end{center}
\end{figure}

\begin{figure}[tbp]
\begin{center}
\parbox{.8\textwidth}{
\epsfxsize=\hsize \epsfbox{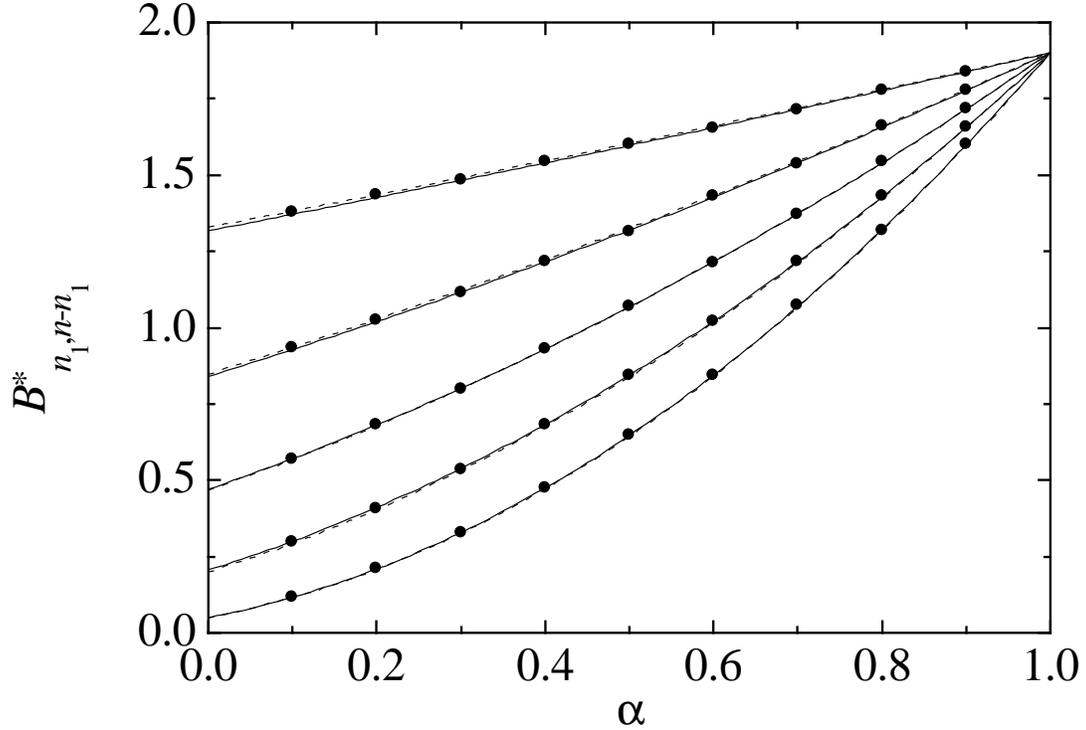}}
\caption{Reduced composition-independent virial coefficients $
B_{n_1,n-n_1}^* $ as functions of the size ratio $\protect\alpha$ for $d=2$, 
$n=6$, and, from top to bottom, $n_1=1$, 2, 3, 4 and 5. The circles are
``exact'' Monte Carlo results \protect\cite{W99a}, the solid line is the
theoretical prediction (\ref{47}) and the dashed line is the theoretical
prediction (\ref{42}). }
\label{fig4}
\end{center}
\end{figure}

\begin{figure}[tbp]
\begin{center}
\parbox{.8\textwidth}{
\epsfxsize=\hsize \epsfbox{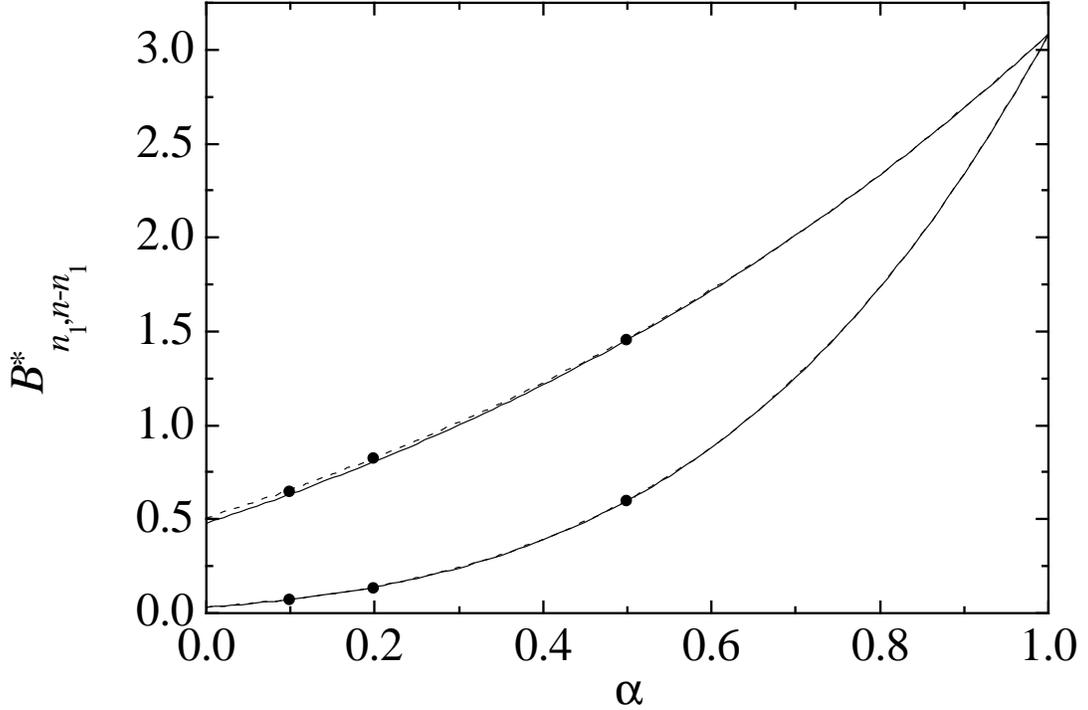}}
\caption{Reduced composition-independent virial coefficients $
B_{n_1,n-n_1}^* $ as functions of the size ratio $\protect\alpha$ for $d=4$, 
$n=3$, and, from top to bottom, $n_1=1$ and 2. The circles are ``exact''
Monte Carlo results \protect\cite{EAGB00}, the solid line is the theoretical
prediction (\ref{11})--(\ref{11.3}) and the dashed line is the theoretical
prediction given by Eqs.\ (\ref{11}) and (\ref{19})--(\ref{27}). }
\label{fig5}
\end{center}
\end{figure}

\begin{figure}[tbp]
\begin{center}
\parbox{.8\textwidth}{
\epsfxsize=\hsize \epsfbox{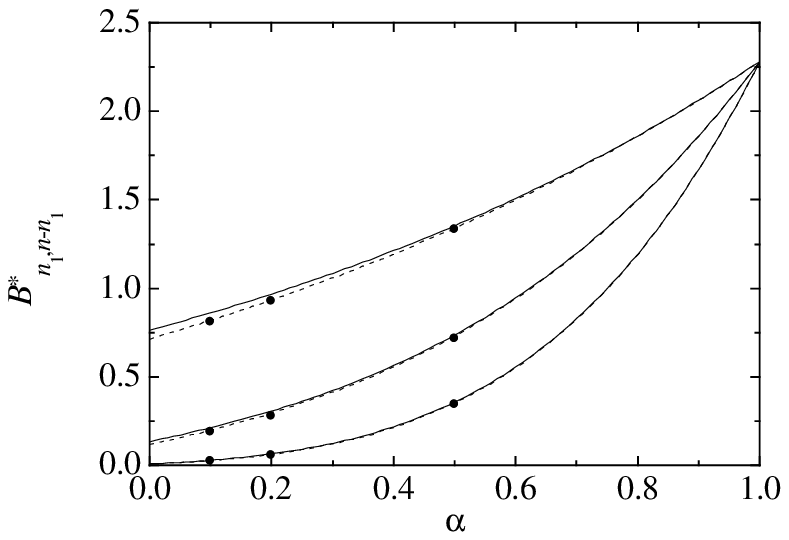}}
\caption{Reduced composition-independent virial coefficients $
B_{n_1,n-n_1}^* $ as functions of the size ratio $\protect\alpha$ for $d=4$, 
$n=4$, and, from top to bottom, $n_1=1$, 2 and 3. The circles are ``exact''
Monte Carlo results \protect\cite{EAGB00}, the solid line is the theoretical
prediction (\ref{11})--(\ref{11.3}) and the dashed line is the theoretical
prediction given by Eqs.\ (\ref{11}) and (\ref{19})--(\ref{27}). }
\label{fig6}
\end{center}
\end{figure}

\begin{figure}[tbp]
\begin{center}
\parbox{.8\textwidth}{
\epsfxsize=\hsize \epsfbox{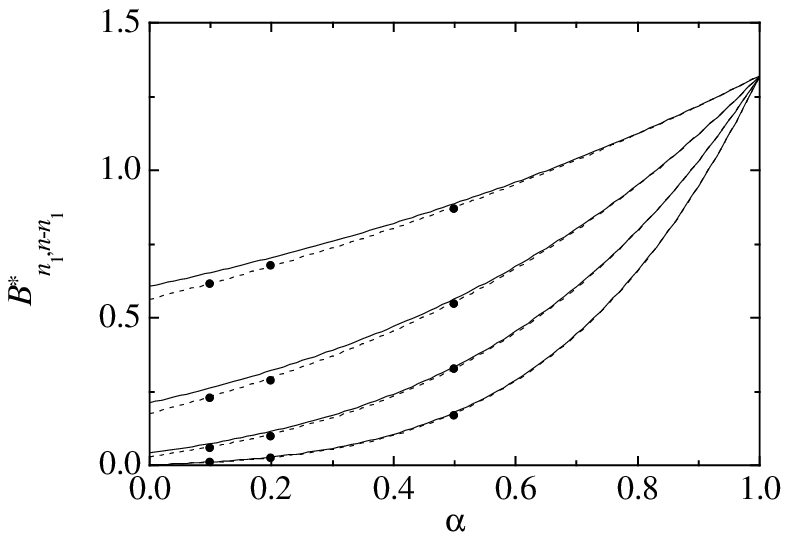}}
\caption{Reduced composition-independent virial coefficients $
B_{n_1,n-n_1}^* $ as functions of the size ratio $\protect\alpha$ for $d=4$, 
$n=5$, and, from top to bottom, $n_1=1$, 2, 3 and 4. The circles are
``exact'' Monte Carlo results \protect\cite{EAGB00}, the solid line is the
theoretical prediction (\ref{11})--(\ref{11.3}) and the dashed line is the
theoretical prediction given by Eqs.\ (\ref{11}) and (\ref{19})--(\ref{27}). 
}
\label{fig7}
\end{center}
\end{figure}

\begin{figure}[tbp]
\begin{center}
\parbox{.8\textwidth}{
\epsfxsize=\hsize \epsfbox{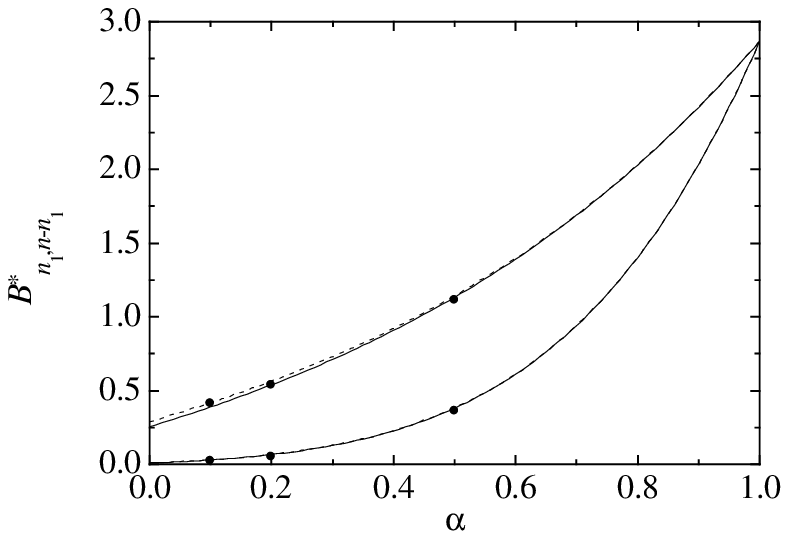}}
\caption{Reduced composition-independent virial coefficients $
B_{n_1,n-n_1}^* $ as functions of the size ratio $\protect\alpha$ for $d=5$, 
$n=3$, and, from top to bottom, $n_1=1$ and 2. The circles are ``exact''
Monte Carlo results \protect\cite{EAGB00}, the solid line is the theoretical
prediction (\ref{11})--(\ref{11.3}) and the dashed line is the theoretical
prediction given by Eqs.\ (\ref{11}) and (\ref{19})--(\ref{27}). }
\label{fig8}
\end{center}
\end{figure}

\begin{figure}[tbp]
\begin{center}
\parbox{.8\textwidth}{
\epsfxsize=\hsize \epsfbox{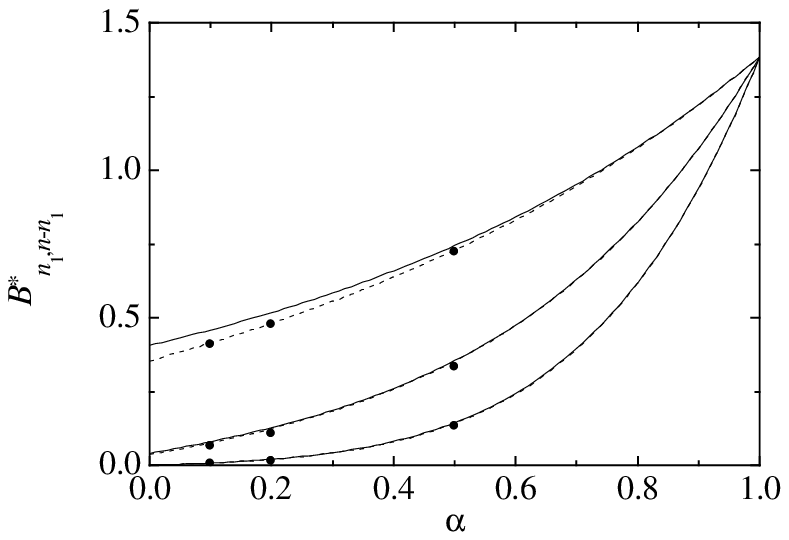}}
\caption{Reduced composition-independent virial coefficients $
B_{n_1,n-n_1}^* $ as functions of the size ratio $\protect\alpha$ for $d=5$, 
$n=4$, and, from top to bottom, $n_1=1$, 2 and 3. The circles are ``exact''
Monte Carlo results \protect\cite{EAGB00}, the solid line is the theoretical
prediction (\ref{11})--(\ref{11.3}) and the dashed line is the theoretical
prediction given by Eqs.\ (\ref{11}) and (\ref{19})--(\ref{27}). }
\label{fig9}
\end{center}
\end{figure}

\begin{figure}[tbp]
\begin{center}
\parbox{.8\textwidth}{
\epsfxsize=\hsize \epsfbox{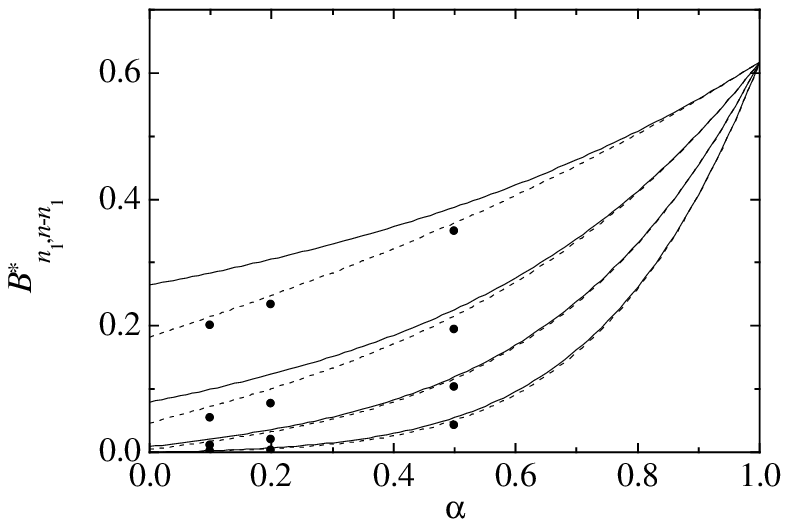}}
\caption{Reduced composition-independent virial coefficients $
B_{n_1,n-n_1}^* $ as functions of the size ratio $\protect\alpha$ for $d=5$, 
$n=5$, and, from top to bottom, $n_1=1$, 2, 3 and 4. The circles are
``exact'' Monte Carlo results \protect\cite{EAGB00}, the solid line is the
theoretical prediction (\ref{11})--(\ref{11.3}) and the dashed line is the
theoretical prediction given by Eqs.\ (\ref{11}) and (\ref{19})--(\ref{27}). 
}
\label{fig10}
\end{center}
\end{figure}

In the case of hard hyperspheres in $d=4$ and $d=5$, Figs.\ 
\ref{fig5}--\ref{fig10} illustrate the performance of Eq.\ (\ref{11}) together 
with (\ref{11.1})--(\ref{11.4}) and of Eqs.\ (\ref{11}) and 
(\ref{19})--(\ref{27})
with respect to the ``exact'' data of Ref.\ \cite{EAGB00}. In these
dimensionalities, it is clear that while the modified version of the
estimates for the virial coefficients continues to exhibit a very good
performance, the use of Eqs.\ (\ref{11})--(\ref{11.4}) presents limitations
for $n=4$ and especially for $n=5$. Nevertheless, the EOS for the binary
mixture derived from Eq.\ (\ref{4.1}) (which yields the original estimates
of the coefficients $B_{n_{1},n_{2}} $), once the proper $Z_{\text{s}}(\eta
) ${ is used in both }$d=4$ and $d=5$, turns out to be very
accurate\cite{AGH}.

Once we have analyzed the behaviour of the virial coefficients, the important
question that now arises is, which is the EOS consistent with the modified
virial coefficients? In the case $d=2$, from Eq.\ (\ref{42}) one gets \cite
{S99} 
\begin{eqnarray}
Z_{\text{m}}(\eta ) &=&\frac{x_{1}}{1-\eta _{2}}Z_{\text{s}}\left( \frac{
\eta _{1}}{1-\eta _{2}}\right) \frac{\sigma _{2}-\sigma _{1}}{\sigma _{2}}+
\frac{x_{2}}{1-\eta _{1}}Z_{\text{s}}\left( \frac{\eta _{2}}{1-\eta _{1}}
\right) \frac{\sigma _{1}-\sigma _{2}}{\sigma _{1}}  \nonumber \\
&&+Z_{\text{s}}(\eta )\frac{\langle \sigma ^{2}\rangle }{\sigma _{1}\sigma
_{2}}.  \label{49}
\end{eqnarray}
The EOS consistent with the coefficients (\ref{19})--(\ref{27}) can also be
explicitly obtained for $d\geq 3$ by carrying out the resummation of the
corresponding virial series. The result is (see Appendix \ref{appB} for
details) 
\begin{eqnarray}
Z_{\text{m}}(\eta ) &=&1+\frac{K}{\langle \sigma ^{d}\rangle }[Z_{\text{s}
}(\eta )-1]+\frac{K_{0}}{\langle \sigma ^{d}\rangle }x_{1}x_{2}\frac{\eta }{
1-\eta }  \nonumber \\
&&+\frac{K_{1}}{\sigma _{2}^{d}}x_{1}\left[ \frac{1}{1-\eta _{2}}Z_{\text{s}
}\left( \frac{\eta _{1}}{1-\eta _{2}}\right) -1\right] +\frac{K_{2}}{\sigma
_{1}^{d}}x_{2}\left[ \frac{1}{1-\eta _{1}}Z_{\text{s}}\left( \frac{\eta _{2}
}{1-\eta _{1}}\right) -1\right] ,  \label{28}
\end{eqnarray}
where $K$, $K_{0}$, $K_{1}$ and $K_{2}$ are independent of the density and
are given by 
\begin{equation}
K=x_{1}^{2}\sigma _{1}^{d}\left( 1-\frac{K_{1}}{\sigma _{2}^{d}}\right)
+x_{2}^{2}\sigma _{2}^{d}\left( 1-\frac{K_{2}}{\sigma _{1}^{d}}\right)
+2x_{1}x_{2}\left( \sigma _{12}^{d}-\frac{K_{1}+K_{2}+K_{0}}{2^{d}}\right) ,
\label{28.1}
\end{equation}
\begin{equation}
K_{0}=\frac{1}{2^{2-d}+d-3}\left[ \sigma _{1}^{d}+\sigma _{2}^{d}+\left(
2^{d-1}-1\right) \sigma _{1}\sigma _{2}\left( \sigma _{1}^{d-2}+\sigma
_{2}^{d-2}\right) -2^{d}\sigma _{12}^{d}\right] ,  \label{28.2}
\end{equation}
\begin{eqnarray}
K_{1} &=&\frac{1}{(2^{d}-4)(2^{2-d}+d-3)\sigma _{12}}\left\{ \sigma _{12}^{d}
\left[ 2^{d}(d-2)\sigma _{1}+8(1-2^{d-1}+d2^{d-3})\sigma _{2}\right] \right.
\nonumber \\
&&-(2^{d-1}-2)\sigma _{1}\sigma _{2}\left( \sigma _{1}^{d-1}+\sigma
_{2}^{d-1}\right) -(d-2)(2^{d-1}-1)\sigma _{1}^{2}\sigma
_{2}^{d-1}-(2^{d-1}-d)\sigma _{1}^{d-1}\sigma _{2}^{2}  \nonumber \\
&&\left. -(d-2)\sigma _{1}^{d+1}-\left[ d-4-2^{d-1}(d-3)\right] \sigma
_{2}^{d+1}\right\} ,  \label{28.3}
\end{eqnarray}
\begin{eqnarray}
K_{2} &=&\frac{1}{(2^{d}-4)(2^{2-d}+d-3)\sigma _{12}}\left\{ \sigma _{12}^{d}
\left[ 2^{d}(d-2)\sigma _{2}-8(1-2^{d-1}+d2^{d-3})\sigma _{1}\right] \right.
\nonumber \\
&&-(2^{d-1}-2)\sigma _{1}\sigma _{2}\left( \sigma _{1}^{d-1}+\sigma
_{2}^{d-1}\right) -(d-2)(2^{d-1}-1)\sigma _{2}^{2}\sigma
_{1}^{d-1}-(2^{d-1}-d)\sigma _{2}^{d-1}\sigma _{1}^{2}  \nonumber \\
&&\left. -(d-2)\sigma _{2}^{d+1}-\left[ d-4-2^{d-1}(d-3)\right] \sigma
_{1}^{d+1}\right\} .  \label{28.4}
\end{eqnarray}
{ Here, $\sigma _{12}\equiv (\sigma_1+\sigma_2)/2$.} Note that Eq.\ 
(\ref{49}) for hard discs is included in the structure of
Eq.\ (\ref{28}), except that then Eqs.\ (\ref{28.1})--(\ref{28.4}) cannot be
applied and we have instead $K_{0}=0$, $K_{1}=\sigma _{2}(\sigma _{2}-\sigma
_{1})$, $K_{2}=\sigma _{1}(\sigma _{1}-\sigma _{2})$ and $K=\langle \sigma
^{2}\rangle \left( 1-x_{1}K_{1}/\sigma _{2}^{2}-x_{2}K_{2}/\sigma
_{1}^{2}\right) $.

Equation (\ref{28}) complies with the exact results given in Eq.\ (\ref{13}). It 
is clearly { more complicated} than the original recipe, Eq.\ (\ref
{4.1}), and should in principle be useful in particular for mixtures
involving components of disparate sizes. In contrast to Eq.\ (\ref{4.1}),
the proposal (\ref{28}) expresses $Z_{\text{m}}(\eta )$ in terms not only of 
$Z_{\text{s}}(\eta )$ but also involves $Z_{\text{s}}\left( \frac{\eta _{1}}{
1-\eta _{2}}\right) $ and $Z_{\text{s}}\left( \frac{\eta _{2}}{1-\eta _{1}}
\right) $.

We note that if $d=3$, it follows from Eqs. (\ref{28.1})--(\ref{28.4}) that $
K_{0}=0$, $K_{1}=\sigma _{2}(\sigma _{1}-\sigma _{2})^{2}$, $K_{2}=\sigma
_{1}(\sigma _{1}-\sigma _{2})^{2}$ and $K=\langle \sigma ^{3}\rangle \left(
1-x_{1}K_{1}/\sigma _{2}^{3}-x_{2}K_{2}/\sigma _{1}^{3}\right) $.
Consequently, Eq.\ (\ref{28}) for hard spheres becomes 
\begin{eqnarray}
Z_{\text{m}}(\eta ) &=&\frac{x_{1}}{1-\eta _{2}}Z_{\text{s}}\left( \frac{
\eta _{1}}{1-\eta _{2}}\right) \frac{(\sigma _{1}-\sigma _{2})^{2}}{\sigma
_{2}^{2}}+\frac{x_{2}}{1-\eta _{1}}Z_{\text{s}}\left( \frac{\eta _{2}}{
1-\eta _{1}}\right) \frac{(\sigma _{1}-\sigma _{2})^{2}}{\sigma _{1}^{2}} 
\nonumber \\
&&+Z_{\text{s}}(\eta )\frac{2\langle \sigma ^{2}\rangle \sigma _{1}\sigma
_{2}-\langle \sigma ^{4}\rangle }{\sigma _{1}^{2}\sigma _{2}^{2}}.
\label{40}
\end{eqnarray}
It is interesting to note that Eqs.\ (\ref{49}) and (\ref{40}) can be
written in the common form 
\begin{eqnarray}
Z_{\text{m}}(\eta ) &=&Z_{\text{s}}(\eta )+x_{1}\left[ \frac{1}{1-\eta _{2}}
Z_{\text{s}}\left( \frac{\eta _{1}}{1-\eta _{2}}\right) -Z_{\text{s}}(\eta )
\right] \left( \frac{\sigma _{2}-\sigma _{1}}{\sigma _{2}}\right) ^{d-1} 
\nonumber \\
&&+x_{2}\left[ \frac{1}{1-\eta _{1}}Z_{\text{s}}\left( \frac{\eta _{2}}{
1-\eta _{1}}\right) -Z_{\text{s}}(\eta )\right] \left( \frac{\sigma
_{1}-\sigma _{2}}{\sigma _{1}}\right) ^{d-1}.  \label{50}
\end{eqnarray}
Equation (\ref{50}) reduces to Eq.\ (\ref{49}) if $d=2$ and to Eq.\ (\ref{40}) 
if $d=3$. In addition, it becomes exact for $d=1$, i.e., $Z_{\text{s}
}(\eta )=Z_{\text{m}}(\eta )=(1-\eta )^{-1}$. It seems then very tempting to
assume Eq.\ (\ref{50}) for arbitrary $d$. With such an assumption, after
using Eqs.\ (\ref{35}), (\ref{36}), (\ref{33}) and (\ref{34}), we simply get
for the composition-independent virial coefficients 
\begin{equation}
B_{n_{1},n_{2}}^{\ast }=v_{d}^{n-1}\left[ b_{n}\frac{n_{1}\alpha ^{d}+n_{2}}{
n}+\left( b_{n_{1}}-b_{n}\right) \frac{n_{1}}{n}\alpha (\alpha
-1)^{d-1}+\left( b_{n_{2}}-b_{n}\right) \frac{n_{2}}{n}(1-\alpha )^{d-1}
\right] .  \label{51}
\end{equation}
Of course, this includes the cases $d=2$ [{\it cf.}\ Eq.\ (\ref{42})] and $
d=3$ [{\it cf.}\ Eq.\ (\ref{43})]. However, Eq.\ (\ref{51}) differs from the
prescription (\ref{19})--(\ref{27}) if $d\geq 4$. Although Eq.\ (\ref{51})
is consistent with conditions (\ref{41}), (\ref{21}) [for $d\geq 3$] and 
(\ref{13}), it does not reproduce the exact second virial coefficient 
$B_{1,1}^{\ast }(\alpha )=v_{d}(1+\alpha )^{d}/2$ { (except, of course, for 
$d\leq 3$)}. In fact, we have checked
that the performance of Eqs.\ (\ref{50}) and (\ref{51}) is very poor for $d\geq 
4$, so that they will not be further considered here.

\section{The equation of state of binary hard-sphere mixtures}

\label{sec4} Up to here, although ocasionally we have mentioned explicit
features of hard discs and hard spheres, the developments presented in the
previous Sections apply for general dimensionality $d$. Given their
intrinsic importance, in this Section we focus on the results for the EOS in
three dimensions. In order to set the framework of our discussion, we shall
first recall two other EOS in the literature which share with ours the idea
of obtaining $Z_{\text{m}}(\eta )$ from the knowledge of $Z_{\text{s}}(\eta) $.

The first one is due to Hamad\cite{H95} who in 1994 \cite{H94} derived a
consistency condition involving the derivatives of the contact values of the
radial distribution functions $g_{ij}$ with respect to $\sigma _{k}$
\cite{BS00}.
Imposing this condition, he proposed the following equation of state for
hard-sphere mixtures 
\begin{equation}
Z_{\text{m}}^{\text{H}}(\eta )=Z_{\text{s}}(\eta )+\frac{3\eta }{(1-\eta
)^{3}}\left[ \frac{\langle \sigma ^{2}\rangle ^{3}}{\langle \sigma
^{3}\rangle ^{2}}\eta +\frac{\langle \sigma \rangle \langle \sigma
^{2}\rangle }{\langle \sigma ^{3}\rangle }(1-\eta )-1\right] ,  \label{52}
\end{equation}
where we have used the label H to refer to Hamad's proposal. It must be
noted that Eq.\ (\ref{52}) was not explicitly written in Ref.\ \cite{H95},
where the results are expressed in a rather more involved form. The second
EOS for binary mixtures of hard spheres that we want to consider was
proposed more recently by Barrio and Solana\cite{BS99} (label BS). It reads 
\begin{equation}
Z_{\text{m}}^{\text{BS}}(\eta )=1+\frac{1}{4}(1+\beta \eta )\left( 1+3\frac{
\langle \sigma \rangle \langle \sigma ^{2}\rangle }{\langle \sigma
^{3}\rangle }\right) \left[ Z_{\text{s}}(\eta )-1\right] ,  \label{61}
\end{equation}
where $\beta $ is adjusted as to reproduce the exact third virial
coefficient \cite{KM75}, namely 
\begin{equation}
\begin{array}{l}
B_{3,0}^{\ast }=v_3^210\alpha ^{3},\quad B_{2,1}^{\ast }=v_3^2\left(\frac{1}{
3}+2\alpha +5\alpha ^{2}+\frac{8}{3}\alpha ^{3}\right), \\ 
B_{1,2}^{\ast }=v_3^2\left(\frac{8}{3}+5\alpha +2\alpha ^{2}+\frac{1}{3}
\alpha ^{3}\right),\quad B_{0,3}^{\ast }=v_3^210.
\end{array}
\label{62}
\end{equation}
More explicitly, in our notation $\beta $ is written as 
\begin{equation}
\beta =\frac{B_{3}}{v_{3}^{2}\langle \sigma ^{3}\rangle ^{2}(1+3\langle
\sigma \rangle \langle \sigma ^{2}\rangle /\langle \sigma ^{3}\rangle )}-
\frac{10}{4}.  \label{65}
\end{equation}

We now first concentrate on the composition-independent virial coefficients.
Our original recipe, Eqs.\ (\ref{11})--(\ref{11.3}), gives for $d=3$ 
\begin{eqnarray}
{B_{n_{1},n_{2}}^{\ast {\text{SYH}}}} &=&v_3^{n-1}\frac{n_{1}n_{2}}{
2n(n-1)(n-2)}\left\{ 2(n_{1}+3n_{2}-4)+\frac{n_{2}-1}{n_{1}}
(n_{1}+2n_{2}-4)b_{n}\right.  \nonumber \\
&&+\left[ 2(n_{1}-5n_{2}+4)+(n_{1}+4n_{2}-5)b_{n}\right] \alpha +\left[
2(n_{2}-5n_{1}+4)+(n_{2}+4n_{1}-5)b_{n}\right] \alpha ^{2}  \nonumber \\
&&\left. +\left[ 2(n_{2}+3n_{1}-4)+\frac{n_{1}-1}{n_{2}}(n_{2}+2n_{1}-4)b_{n}
\right] \alpha ^{3}\right\} ,  \label{48}
\end{eqnarray}
where we have incorporated the label SYH to distinguish it from the other
results. On the other hand, in the three-dimensional case, Eqs.\ 
(\ref{19})--(\ref{27}) yield 
\begin{eqnarray}
{B_{n_{1},n_{2}}^{\ast {\text{W}}}} &=&v_3^{n-1}\left[\frac{n_{2}}{n}
b_{n_{2}}+\left( \frac{2n_{2}-n_{1}}{n}b_{n}+\frac{n_{1}}{n}b_{n_{1}}-2\frac{
n_{2}}{n}b_{n_{2}}\right) \alpha\right.  \nonumber \\
&&\left.+\left( \frac{2n_{1}-n_{2}}{n}b_{n}+\frac{n_{2}}{n}b_{n_{2}}-2\frac{
n_{1}}{n}b_{n_{1}}\right) \alpha ^{2}+\frac{n_{1}}{n}b_{n_{1}}\alpha ^{3}
\right],  \label{43}
\end{eqnarray}
which coincides with the form proposed by Wheatley \cite{W99b} and so the
label in this case is W. The virial coefficients associated with Eqs. (\ref
{52}) and (\ref{61}) are also readily derived. In the former case, first it
is easy to obtain 
\begin{equation}
B_{n}^{\text{H}}=v_{3}^{n-1}\langle \sigma ^{3}\rangle ^{n-1}\left[ b_{n}-
\frac{3}{2}n(n-1)+3(n-1)\frac{\langle \sigma \rangle \langle \sigma
^{2}\rangle }{\langle \sigma ^{3}\rangle }+\frac{3}{2}(n-1)(n-2)\frac{
\langle \sigma ^{2}\rangle ^{3}}{\langle \sigma ^{3}\rangle ^{2}}\right] .
\label{55}
\end{equation}
Now, by making $m=0$ and $m=1$ in Eq.\ (\ref{6}) and $m=1$ in Eq.\ (\ref{5}), 
and inserting the results into Eq.\ (\ref{55}), we get { the 
composition-independent coefficients arising from Hamad's proposal,}
\begin{eqnarray}
{B_{n_{1},n_{2}}^{\ast {\text{H}}}} &=&v_3^{n-1}\left\{\frac{n_{2}}{n}\left[
b_{n}-\frac{3}{2}n_{1}(n-1+n_{2})\right] +\frac{3}{2}\frac{n_{1}n_{2}}{n}
(3n_{2}-1)\alpha \right.  \nonumber \\
&&\left.+\frac{3}{2}\frac{n_{1}n_{2}}{n}(3n_{1}-1)\alpha ^{2}+\frac{n_{1}}{n}
\left[ b_{n}-\frac{3}{2}n_{2}(n-1+n_{1})\right] \alpha ^{3}\right\}.
\label{59}
\end{eqnarray}
As far as Eq.\ (\ref{61}) is concerned, one readily derives (for $n\geq 3$) 
\begin{eqnarray}
B_{n}^{\text{BS}} &=&\frac{v_{3}^{n-1}}{4}(b_{n}+\beta b_{n-1})\langle
\sigma ^{3}\rangle ^{n-1}\left( 1+3\frac{\langle \sigma \rangle \langle
\sigma ^{2}\rangle }{\langle \sigma ^{3}\rangle }\right)  \nonumber \\
&=&\frac{v_{3}^{n-1}}{4}\left( b_{n}-\frac{10}{4}b_{n-1}\right) \langle
\sigma ^{3}\rangle ^{n-1}\left( 1+3\frac{\langle \sigma \rangle \langle
\sigma ^{2}\rangle }{\langle \sigma ^{3}\rangle }\right)  
+\frac{v_{3}^{n-3}}{4}b_{n-1}B_{3}\langle \sigma ^{3}\rangle ^{n-3}.
\label{63}
\end{eqnarray}
Now, after some algebra also involving the substitution of the results
obtained by taking $m=0$ and $m=1$ in Eq.\ (\ref{6}) into Eq.\ (\ref{63}),
the composition-independent virial coefficients for the Barrio-Solana EOS
turn out to be given by 
\begin{eqnarray}
{B_{n_{1},n_{2}}^{\ast {\text{BS}}}} &=&v_3^{n-1}\frac{b_{n}}{4n(n-1)}\left[
n_{2}(3n_{2}+n-4)+3n_{1}n_{2}\alpha +3n_{1}n_{2}\alpha
^{2}+n_{1}(3n_{1}+n-4)\alpha ^{3}\right]  \nonumber \\
&&-v_3^{n-1}\frac{3b_{n-1}n_{1}n_{2}}{8n(n-1)(n-2)}\left[
n+2n_{2}-4+(n-6n_{2}+4)\alpha \right.  \nonumber \\
&&\left. +(n-6n_{1}+4)\alpha ^{2}+(n+2n_{1}-4)\alpha ^{3}\right] .
\label{64}
\end{eqnarray}

These four approximations for $B_{n_{1},n_{2}}^{\ast }$ satisfy the
requirements (\ref{41}) and (\ref{21}). The condition (\ref{13}) is only
satisfied by the W approximation, while the condition (\ref{13bis}) is only
satisfied by the H approximation. By construction, the BS approximation,
Eq.\ (\ref{64}), reproduces the exact\/ third virial coefficients (\ref{62}). It 
is interesting to point out the approximations (\ref{48}), (\ref{43})
and (\ref{59}) also yield the exact value of the third virial coefficients.

In order to assess the accuracy of the previous approximate expressions for $
B_{n_{1},n_{2}}^{\ast }$ with $n\geq 4$, we will take the case $n_{1}=3$, $
n_{2}=1$ as a benchmark for comparison since the coefficient $B_{3,1}^{\ast
} $ has been obtained {\em exactly\/} for the interval $0\leq \alpha \leq 2/
\sqrt{3}-1\simeq 0.1547$ \cite{B98}. This exact result is 
\begin{eqnarray}
B_{3,1}^{\ast } &=&\frac{v_3^{3}}{4}\left( 1+9\alpha +36\alpha ^{2}+21\alpha
^{3}+\frac{27}{2}\alpha ^{4}+\frac{27}{10}\alpha ^{5}\right.  \nonumber \\
&&\left. -\frac{108}{5}\alpha ^{6}-\frac{648}{35}\alpha ^{7}-\frac{81}{14}
\alpha ^{8}-\frac{9}{14}\alpha ^{9}\right) .  \label{46}
\end{eqnarray}
Setting $n_{1}=3$ and $n_{2}=1$ in Eqs.\ (\ref{48}), (\ref{43}), (\ref{59})
and (\ref{64}), respectively, yields 
\begin{equation}
{B_{3,1}^{\ast {\text{SYH}}}}=\frac{v_{3}^{3}}{4}\left[ 1+\left( \frac{b_{4}
}{2}+1\right) \alpha +\left( 2b_{4}-5\right) \alpha ^{2}+3\left( \frac{b_{4}
}{2}+1\right) \alpha ^{3}\right] ,  \label{45}
\end{equation}
\begin{equation}
{B_{3,1}^{\ast {\text{W}}}}=\frac{v_{3}^{3}}{4}\left[ 1+\left(
28-b_{4}\right) \alpha +\left( 5b_{4}-59\right) \alpha ^{2}+30\alpha ^{3}
\right] ,  \label{44}
\end{equation}
\begin{equation}
{B_{3,1}^{\ast {\text{H}}}}=\frac{v_{3}^{3}}{4}\left[ b_{4}-18+9\alpha
+36\alpha ^{2}+3(b_{4}-9)\alpha ^{3}\right]  \label{60}
\end{equation}
and 
\begin{equation}
{B_{3,1}^{\ast {\text{BS}}}}=\frac{v_{3}^{3}}{16}\left[ b_{4}-15+(3b_{4}-15)
\alpha +(3b_{4}+75)\alpha ^{2}+(9b_{4}-45)\alpha ^{3}\right] .  \label{66}
\end{equation}
Note that Eqs.\ (\ref{45}) and (\ref{44}) would coincide if $b_{4}$ were
equal to $18$, as happens with the CS equation of state. However, they
differ if we take the exact value $b_{4}=18.36477$. The maximum deviation of
(\ref{44}) from the exact result (\ref{46}) is about $1.8\%$ (at $\alpha
\simeq 0.08$), while that of (\ref{45}) is about $3.7\%$ (at $\alpha \simeq
0.09$). Also, if $b_{4}$ were equal to $19$, Eq.\ (\ref{60}) would coincide
with Eq.\ (\ref{44}), while using the exact value for $\ b_{4}$ Eq.\ (\ref
{60}) implies that $\lim_{\alpha \rightarrow 0}{B_{3,1}^{\ast {\text{H}}}}
=0.091v_{3}^{3}$ instead of the exact value $\lim_{\alpha \rightarrow
0}B_{3,1}^{\ast }=\frac{1}{4}v_{3}^{3}$, which, however, is satisfied by the
approximations (\ref{45}) and (\ref{44}). Comparison between Hamad's
approximation (\ref{60}) and the exact result (\ref{46}) for $\alpha \leq 2/
\sqrt{3}-1$ shows that the relative error of (\ref{60}) monotonically
decreases from about $64\%$ at $\alpha =0$ to about $18\%$ at $\alpha =2/
\sqrt{3}-1$. Finally, according to Eq. (\ref{66}) with the exact value of $
b_{4}$, $\lim_{\alpha \rightarrow 0}{B_{3,1}^{\ast {\text{BS}}}}
=0.210v_{3}^{3}$, while in the interval $\alpha \leq 2/\sqrt{3}-1$ the
relative error of (\ref{66}) monotonically decreases from about $16\%$ at $
\alpha =0$ to about $1.7\%$ at $\alpha =2/\sqrt{3}-1$. This relative error
is always larger than that of Eqs.\ (\ref{45}) and (\ref{44}), but smaller
than that of Eq.\ (\ref{60}).

\begin{figure}[tbp]
\begin{center}
\parbox{.8\textwidth}{
\epsfxsize=\hsize \epsfbox{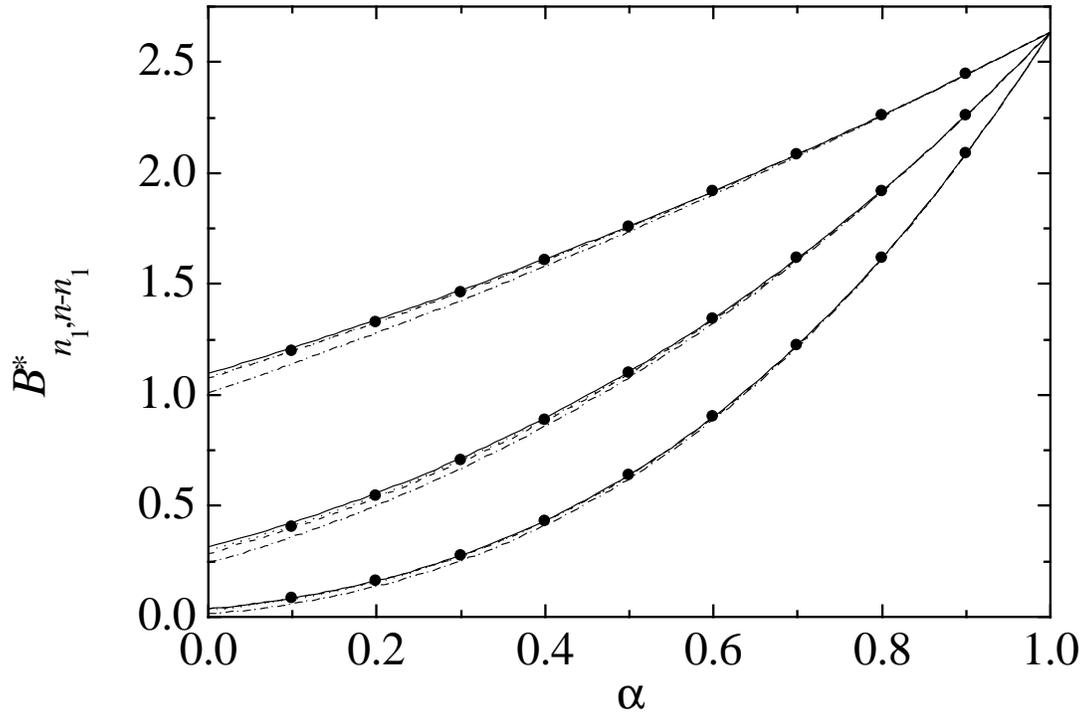}}
\caption{Reduced composition-independent virial coefficients $
B_{n_1,n-n_1}^* $ as functions of the size ratio $\protect\alpha$ for $d=3$, 
$n=4$, and, from top to bottom, $n_1=1$, 2, and 3. The circles are ``exact''
Monte Carlo results \protect\cite{SFG97}, the solid line is the theoretical
prediction (\ref{48}), the dashed line is the theoretical prediction (\ref
{43}), the dash-dot line is the theoretical prediction (\ref{59}) and the
dotted line is the theoretical prediction (\ref{64}). }
\label{fig11}
\end{center}
\end{figure}

\begin{figure}[tbp]
\begin{center}
\parbox{.8\textwidth}{
\epsfxsize=\hsize \epsfbox{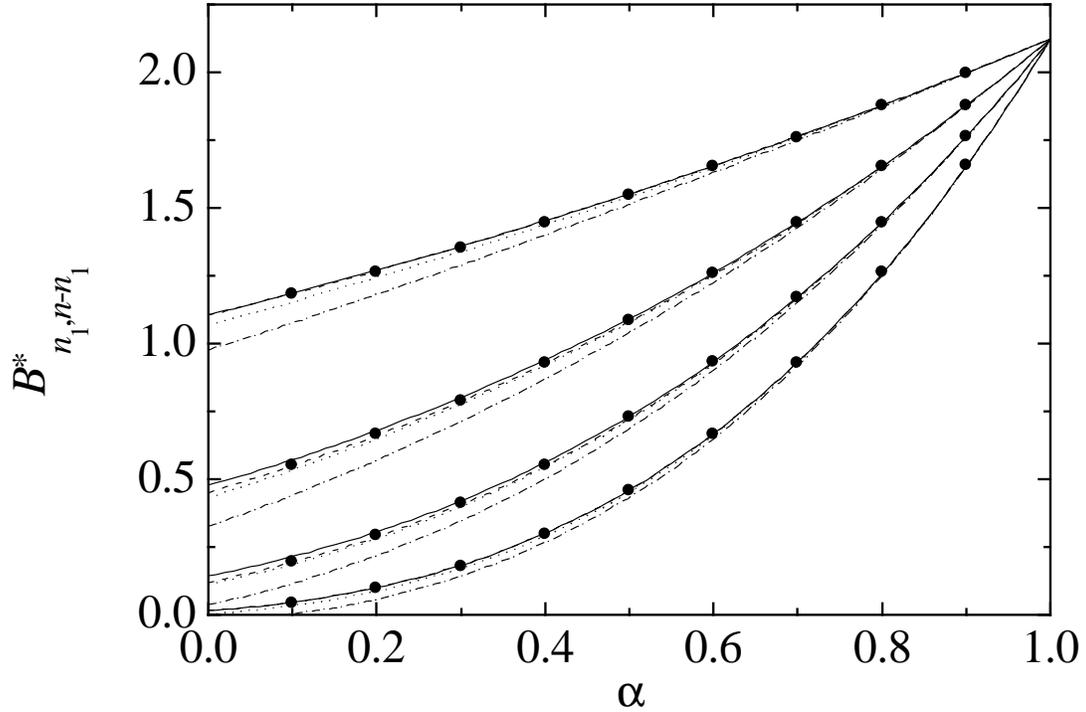}}
\caption{Reduced composition-independent virial coefficients $
B_{n_1,n-n_1}^* $ as functions of the size ratio $\protect\alpha$ for $d=3$, 
$n=5$, and, from top to bottom, $n_1=1$, 2, 3 and 4. The circles are
``exact'' Monte Carlo results \protect\cite{WSG98,EAGB98}, the solid line is
the theoretical prediction (\ref{48}), the dashed line is the theoretical
prediction (\ref{43}), the dash-dot line is the theoretical prediction (\ref
{59}) and the dotted line is the theoretical prediction (\ref{64}). }
\label{fig12}
\end{center}
\end{figure}

Apart from the exact result for ${B}_{3,1}^*$, as pointed
out before, numerical values of the coefficients ${B}_{n_{1},n_{2}}^*$ 
corresponding to $n\leq 5$ are also
available for hard spheres\cite{EACG97,SFG97,EAGB98,WSG98}. In Figs.\ 
\ref{fig11} and \ref{fig12} we display the comparison between the theoretical
estimates we have just discussed for the virial coefficients and the
numerical results. 
{ It is clear that the estimates $B_{n_1,n_2}^{*\text{SYH}}$, 
$B_{n_1,n_2}^{*\text{W}}$ and $B_{n_1,n_2}^{*\text{BS}}$
for $n=4$ and $n=5$ are remarkably close to one another and to the exact
results, except perhaps in the case of $B_{1,4}^{*\text{BS}}$. On the other 
hand, the prediction of $B_{n_1,n_2}^{*\text{H}}$ is much poorer.}

Once we have briefly discussed the performance of the various approximate
expressions for the composition-independent virial coefficients, we may next
wonder how well the approximate EOS for a binary mixture of hard spheres
perform with respect to the simulation data. To address this issue, we have
first to specify what $Z_{\text{s}}(\eta)$ is. Two choices will be made
here. On the one hand, since perhaps the most popular and widely known EOS
for a single component fluid of hard spheres is the CS EOS, we will also
use it here. On the other hand, a more accurate EOS for the same system is
obtained if a rescaled Pad\'{e} approximation taking into account the known
virial coefficients is used\cite{MV99,note}. In Tables \ref{table3} and 
\ref{table4} we present a comparison of the results obtained using the various
approximate EOS for the mixture with these two choices of $Z_{\text{s}}(\eta)$, 
two different { mole} fractions and a diameter ratio $\alpha =0.3$. In Table 
\ref{table3}, the values corresponding to the widely used
Boubl\'{\i}k-Mansoori-Carnahan-Starling-Leland (BMCSL) EOS \cite{BMCSL} have 
also
been included. One immediately notes that, as already stated in the
literature\cite{MV99,SYH99,CCHW00}, the EOS given in Eq.\ (\ref{4.1}) using
the CS EOS for the single component system provides a very accurate account
of the simulation data, better than any of the other EOS for mixtures
considered, including that of Eq.\ (\ref{40}). If instead of the CS EOS the
rescaled Pad\'{e} approximation is taken for $Z_{\text{s}}(\eta )$, then it
is Eq.\ (\ref{40}) which gives { in general the best results, although the 
original recipe
(\ref{4.1}) and the BS EOS (\ref{61}) are not too far behind, the latter being 
especially accurate in the case  $x_1=0.75$.}
It should be 
noted that, irrespective
of the choice for $Z_{\text{s}}(\eta )$, both $Z_{\text{m}}^{\text{H}}(\eta)$ 
and $Z_{\text{m}}^{\text{BS}}(\eta)$ for these size ratio and packing
fractions always underestimate the simulation results, a comment that also
applies to the BMCSL EOS.

\begin{table}[tbp]
\caption{{ Comparison of the compressibility factor from several equations of 
state with simulation values ($Z_{\text{simul}}$) for binary additive mixtures 
of hard spheres
($d=3$) with a size ratio $\alpha=0.3$. The fourth to seventh columns give the 
deviations from  $Z_{\text{simul}}$ of the values of the compressibility factor 
as obtained from the equations of
state (\protect\ref{4.1}), (\protect\ref{40}), (\protect\ref{52}) and 
(\protect\ref{61}), respectively. In these equations of state, the
Carnahan-Starling  compressibily factor for the one-component
fluid \protect\cite{CS69} has been used. For the sake of comparison, also the 
results arising from 
the Boubl\'{\i}k-Mansoori-Carnahan-Starling-Leland (BMCSL) equation of state 
\protect\cite{BMCSL} are included.}\label{table3}}
\begin{tabular}{llrrrrrr}
$x_1$&$\eta$ & $Z_{\text{simul}}$\tablenote{Ref.\ \protect\cite{BMLR96}}&Eq.\ 
(\protect\ref{4.1})\tablenote{Ref.\ \protect\cite{SYH99}}& 
Eq.\ (\protect\ref{40})\tablenote{This work}& Eq.\ 
(\protect\ref{52})\tablenote{Ref.\ \protect\cite{H95}}& Eq.\ 
(\protect\ref{61})\tablenote{Ref.\ \protect\cite{BS99}}& BMCSL\\
\tableline
0.0625&
0.3 & 2.790&$-0.001$ & $-0.009$ & $-0.061$ & $-0.016$& $-0.014$\\
 &0.35 &3.473& 0.006 &$-0.011$ & $-0.112$ & $-0.024$ &$-0.020$\\
  &   0.4 &4.410& 0.013 & $-0.019$ & $-0.210$ & $-0.043$ & $-0.035$\\
   &  0.45 &5.722& 0.027 & $-0.033$ & $-0.385$ & $-0.076$ &$-0.063$\\
   &0.49&7.158&0.065 & $-0.036$ & $-0.603$ & $-0.102$ & $-0.081$\\
0.75&0.3 & 3.554&$-0.005$ & $-0.006$ & $-0.024$ &$ -0.010$ &$-0.008$\\
&0.35 & 4.601&$-0.012$ &$-0.014$& $-0.048$ & $-0.021$ &$-0.018$\\
&0.4 & 6.045&$-0.010$ & $-0.014$ & $-0.079$ &$ -0.027$ &$ -0.021$\\
& 0.45 &8.097&$ -0.002$ & $-0.011$ & $-0.129$ &$-0.033$ & $-0.022$\\
 & 0.49 &10.415&$-0.004$ & $-0.021$ & $-0.210$ &$-0.056$ & $-0.037$
\end{tabular}
\end{table}

\begin{table}[tbp]
\caption{{ Comparison of the compressibility factor from several equations of 
state with simulation values ($Z_{\text{simul}}$) for binary additive mixtures 
of hard spheres
($d=3$) with a size ratio $\alpha=0.3$. The fourth to seventh columns give the 
deviations from  $Z_{\text{simul}}$ of the values of the compressibility factor 
as obtained from the equations of
state (\protect\ref{4.1}), (\protect\ref{40}), (\protect\ref{52}) and 
(\protect\ref{61}), respectively.
 In these equations of state, a
rescaled Pad\'e approximant compressibily factor for the one-component
fluid \protect\cite{MV99,note} has been used.}\label{table4}}
\begin{tabular}{llrrrrr}
$x_1$&$\eta$ &  $Z_{\text{simul}}$\tablenote{Ref.\ \protect\cite{BMLR96}}&Eq.\ 
(\protect\ref{4.1})\tablenote{Ref.\ \protect\cite{SYH99}}& 
Eq.\ (\protect\ref{40})\tablenote{This work}& Eq.\ 
(\protect\ref{52})\tablenote{Ref.\ \protect\cite{H95}}& Eq.\ 
(\protect\ref{61})\tablenote{Ref.\ \protect\cite{BS99}}\\
\tableline
0.0625&
0.3 & 2.790&0.004 & $-0.005$ & $-0.050$ & $-0.010$  \\
&0.35 & 3.473&0.014 & $-0.004$ & $-0.095$ & $-0.014$  \\
 &    0.4 &4.410& 0.025 & $-0.008$ & $-0.186$ & $-0.029$  \\
  &    0.45 &5.722& 0.043 & $-0.018$ & $-0.353$ & $-0.058$  \\
   &    0.49 &7.158& 0.084 & $-0.015$ & $-0.564$ & $-0.081$ \\
0.75& 0.3 & 3.554&0.003 & 0.002 & $-0.013$ & $-0.001$  \\
     &0.35 &4.601& 0.001 & $-0.001$ & $-0.032$ &$-0.007$ \\
&     0.4 &6.045& 0.010  & 0.006 &$ -0.055$ &$-0.007$\\
 &    0.45 & 8.097&0.025 & 0.016 & $-0.096$ &$-0.006$  \\
  &   0.49 &10.415& 0.028 & 0.012 & $-0.171$ &  $-0.023$  \\
 \end{tabular}
\end{table}

\section{ Concluding remarks}

\label{sec5}

In this paper we have further assessed the merits and limitations of a
simple recipe recently introduced\cite{SYH99} to derive the compressibility
factor of a multicomponent mixture of $d$-dimensional hard spheres. We have
now considered the case of binary mixtures and looked in particular at the
composition-independent virial coefficients. The comparison with the
available exact and simulation results confirmed the usefulness of our
approach except perhaps when the mixture involves components of very
disparate sizes { (especially for high dimensionalities)}. Guided by the 
conditions that the virial coefficients must
fulfill in certain limits, a slight modification of the form of the original
coefficients was made, trying to sacrifice simplicity as less as possible
while at the same time improving the accuracy of their numerical
predictions. It is fair to state that the (modified) composition-independent
virial coefficients are in excellent agreement with the reported values for
all dimensions and all size ratios.  

Due to its particular relevance, the case of the binary mixture of hard
spheres $(d=3)$ was analyzed in some detail, both in connection with the
virial coefficients and through the consideration of different EOS for this
system that exist in the literature. As for the virial coefficients, 
the
trends pointed out above { about the reliability of the original and modified 
prescriptions} hold also for $d=3$.  The comparison with
the simulation results for the compressibility factor of these mixtures
given in Tables \ref{table3} and \ref{table4} indicates the superiority of
the original recipe for the EOS provided one uses the CS EOS for 
$Z_{\text{s}}(\eta )$. On the other hand, if the more accurate rescaled Pad\'{e}
approximation for $Z_{\text{s}}(\eta )$ is used, it is the new EOS  
(\ref{40}) (i.e., the one obtained after resumming the virial series with
the modified coefficients) which gives the best performance, although only
slightly better than $Z_{\text{m}}(\eta )$ as given by Eq.\ (\ref{4.1}) with $
d=3$. 
Both equations are also of comparable accuracy to the one of the presumably 
best EOS
available presently for hard-sphere mixtures, namely the rescaled Pad\'e
approximant for mixtures introduced by Malijevsk\'y and Veverka \cite{MV99}.
This latter equation of state rests on a different philosophy, however, in
the sense that it makes use of the known virial coefficients of the mixture
rather than the $Z_s(\eta)$, as in our approach and those of Hamad \cite{H95}
and Barrio and Solana \cite{BS99}.

In conclusion, we want to add that both the original proposal of the EOS of
a binary mixture of $d$-dimensional hard spheres, Eq.\ (\ref{4.1}), and the
modified version, Eq.\ (\ref{28}), provide two simple and yet accurate EOS at
least with respect to the composition-independent virial coefficients and in
comparison with the (relatively scarce) simulation data. Further confirmation
of their usefulness depends on new simulations which we hope will be
encouraged by the results presented in this paper.

\acknowledgments
{ A.S. and S.B.Y. 
acknowledge partial support from the DGES (Spain) through grant No.\
PB97-1501 and from the Junta de Extremadura (Fondo Social Europeo) through
grant No.\ IPR99C031. They are also grateful to the DGES (Spain) for 
sabbatical grants No.\ PR2000-0117 and No.\ PR2000-0116, 
respectively. The research of M.L.H. was
supported in part by DGAPA-UNAM under Project IN103100.}
\appendix

\section{Derivation of Eqs.\ (\ref{11})--(\ref{11.3})}

\label{appA} In order to identify the composition-independent coefficients $
B_{n_1,n_2}$ from Eq.\ (\ref{3}), we need to expand the moments $\langle
\sigma^p\rangle$. After some algebra, we get 
\begin{eqnarray}
\langle \sigma ^{d}\rangle^{n-3} \langle \sigma ^{d-m}\rangle\langle \sigma
^{m+1}\rangle \langle \sigma ^{d-1}\rangle
&=&\sigma_1^{d(n-1)}\sum_{n_1=0}^n \alpha^{d(n-n_1-1)}\frac{(n-3)!}{n_1!
(n-n_1)!}x_1^{n_1}x_2^{n-n_1}  \nonumber \\
&&\times\left[n_1(n_1-1)(n_1-2)\alpha^d+n_1(n_1-1)(n-n_1)\alpha^{d-1} \right.
\nonumber \\
&&+n_1(n_1-1)(n-n_1)\alpha^{d-m} +n_1(n-n_1)(n-n_1-1)\alpha^{d-m-1} 
\nonumber \\
&& +n_1(n_1-1)(n-n_1)\alpha^{m+1}+n_1(n-n_1)(n-n_1-1)\alpha^{m}  \nonumber \\
&& +n_1(n-n_1)(n-n_1-1) \alpha  \nonumber \\
&&\left.+(n-n_1)(n-n_1-1)(n-n_1-2)\right],  \label{5}
\end{eqnarray}
\begin{eqnarray}
\langle \sigma ^{d}\rangle^{n-2} \langle \sigma ^{d-m}\rangle\langle \sigma
^{m}\rangle &=&\sigma_1^{d(n-1)}\sum_{n_1=0}^n \alpha^{d(n-n_1-1)}\frac{
(n-2)!}{n_1! (n-n_1)!}x_1^{n_1}x_2^{n-n_1}  \nonumber \\
&&\times\left[n_1(n_1-1)\alpha^d +n_1(n-n_1)\alpha^{d-m}\right.  \nonumber \\
&&\left. +n_1(n-n_1)\alpha^{m} +(n-n_1)(n-n_1-1)\right],  \label{6}
\end{eqnarray}
where $\alpha\equiv \sigma_2/\sigma_1$. Now we use the identities 
\begin{equation}
\sum_{m=0}^d \frac{d!}{m!(d-m)!}\alpha^m=(1+\alpha)^d,  \label{7}
\end{equation}
\begin{equation}
\sum_{m=0}^d \frac{(d-1)!}{m!(d-m)!}(d-m) \alpha^m=(1+\alpha)^{d-1},
\label{8}
\end{equation}
to get 
\begin{eqnarray}
&&\sum_{m=0}^d \frac{(d-1)!}{m!(d-m)!}(d-m)\langle \sigma ^{d}\rangle^{n-3}
\langle \sigma ^{d-m}\rangle\langle \sigma ^{m+1}\rangle \langle \sigma
^{d-1}\rangle = \sigma_1^{d(n-1)}\sum_{n_1=0}^n \alpha^{d(n-n_1-1)}\frac{
(n-3)!}{n_1! (n-n_1)!}  \nonumber \\
&&\times x_1^{n_1}x_2^{n-n_1} \left[n_1(n_1-1)(n_1-2)2^{d-1}
\alpha^d+n_1(n_1-1)(n-n_1)2^{d-1} \alpha^{d-1} \right.  \nonumber \\
&&
+2n_1(n_1-1)(n-n_1)\alpha(1+\alpha)^{d-1}+2n_1(n-n_1)(n-n_1-1)(1+
\alpha)^{d-1}  \nonumber \\
&&\left. +n_1(n-n_1)(n-n_1-1)2^{d-1} \alpha+(n-n_1)(n-n_1-1)(n-n_1-2)2^{d-1} 
\right],  \label{9}
\end{eqnarray}
\begin{eqnarray}
&&\sum_{m=0}^d \frac{d!}{m!(d-m)!}\langle \sigma ^{d}\rangle^{n-2} \langle
\sigma ^{d-m}\rangle\langle \sigma ^{m}\rangle
=\sigma_1^{d(n-1)}\sum_{n_1=0}^n \alpha^{d(n-n_1-1)}\frac{(n-2)!}{n_1!
(n-n_1)!}x_1^{n_1}x_2^{n-n_1}  \nonumber \\
&&\times\left[n_1(n_1-1)2^d\alpha^d +2n_1(n-n_1)(1+\alpha)^{d}
+(n-n_1)(n-n_1-1)2^d\right].  \label{10}
\end{eqnarray}
{}From here we finally obtain Eqs.\ (\ref{11})--(\ref{11.3}).

\section{Derivation of Eqs.\ (\ref{28})--(\ref{28.4})}
\label{appB} 
Let us rewrite Eq.\ (\ref{11}) as 
\begin{eqnarray}
B_{n_1,n_2}&=&v_d^{n-1} \left[C_{n_1,n_2}^\one
\sigma_1^{d(n_1-1)}\sigma_2^{dn_2}
+C_{n_1,n_2}^\two\sigma_1^{d(n_1-1)+1}\sigma_2^{dn_2-1} +
C_{n_1,n_2}^\three2^{d-1}\sigma_{12}^{d-1}
\sigma_1^{d(n_1-1)+1}\sigma_2^{d(n_2-1)} \right.  \nonumber \\
&&\left. +1\leftrightarrow 2 \right],  \label{31}
\end{eqnarray}
where $\sigma_{12}\equiv (\sigma_1+\sigma_2)/2$
 and the notation $1\leftrightarrow 2$  indicates that a contribution similar
to the three summands appearing in the brackets in which the roles of 
 $(n_1,n_2)$ and of $(\sigma_1,\sigma_2)$ are interchanged (i.e., $n_1\to n_2$, 
 $\sigma_{1}\to \sigma _{2}$  and vice versa) should be added.
The compressibility factor of the mixture is then given by 
\begin{eqnarray}
Z_{\text{m}}(\eta)-1&=&\sum_{n=2}^\infty \rho^{n-1}\sum_{n_1=0}^n B_{n_1,n_2}
\frac{n!}{n_1! n_2!} x_1^{n_1} x_2^{n_2}  \nonumber \\
&=&\sum_{n=2}^\infty (v_d \rho)^{n-1}\sum_{n_1=0}^n\frac{n!}{n_1! n_2!}
x_1^{n_1} x_2^{n_2} \left[C_{n_1,n_2}^\one
\sigma_1^{d(n_1-1)}\sigma_2^{dn_2}
+C_{n_1,n_2}^\two\sigma_1^{d(n_1-1)+1}\sigma_2^{dn_2-1}\right.  \nonumber \\
&&\left.+ C_{n_1,n_2}^\three2^{d-1}\sigma_{12}^{d-1}
\sigma_1^{d(n_1-1)+1}\sigma_2^{d(n_2-1)} \right]  \nonumber \\
&& +1\leftrightarrow 2.  \label{32}
\end{eqnarray}
Now we insert Eqs.\ (\ref{19})--(\ref{27}) and sum over $n_1$ and $n$. To do
that, we use the identities 
\begin{equation}
{x_1}\left[Z_{\text{s}}\left(\eta\right)-1\right]= \sum_{n=2}^\infty (v_d
\rho)^{n-1}\sum_{n_1=0}^n\frac{n!}{n_1! n_2!} x_1^{n_1} x_2^{n_2} b_{n}\frac{
n_1}{n}\sigma_1^{d(n_1-1)}\sigma_2^{dn_2},  \label{35}
\end{equation}
\begin{equation}
{x_2}\left[Z_{\text{s}}\left(\eta\right)-1\right]= \sum_{n=2}^\infty (v_d
\rho)^{n-1}\sum_{n_1=0}^n\frac{n!}{n_1! n_2!} x_1^{n_1} x_2^{n_2} b_{n}\frac{
n_2}{n}\sigma_1^{dn_1}\sigma_2^{d(n_2-1)},  \label{36}
\end{equation}
\begin{equation}
{\langle \sigma^d\rangle}\left[Z_{\text{s}}(\eta)-1\right]=
\sum_{n=2}^\infty (v_d \rho)^{n-1}\sum_{n_1=0}^n\frac{n!}{n_1! n_2!}
x_1^{n_1} x_2^{n_2}b_{n}\sigma_1^{d n_1}\sigma_2^{d n_2},  \label{37}
\end{equation}
\begin{equation}
x_1 x_2\frac{Z_{\text{s}}(\eta)-1}{\langle \sigma^d\rangle}=
\sum_{n=2}^\infty (v_d \rho)^{n-1}\sum_{n_1=0}^n\frac{n!}{n_1! n_2!}
x_1^{n_1} x_2^{n_2} b_{n}\frac{n_1 n_2}{n(n-1)}\sigma_1^{d(n_1-1)}
\sigma_2^{d(n_2-1)},  \label{38}
\end{equation}
\begin{equation}
x_1 x_2 \frac{\eta}{(1-\eta)\langle \sigma^d\rangle}= \sum_{n=2}^\infty (v_d
\rho)^{n-1}\sum_{n_1=0}^n\frac{n!}{n_1! n_2!} x_1^{n_1} x_2^{n_2} \frac{n_1
n_2}{n(n-1)}\sigma_1^{d(n_1-1)}\sigma_2^{d(n_2-1)}.  \label{39}
\end{equation}
In addition,\cite{S99} 
\begin{equation}
\frac{x_1}{1-\eta_2}Z_{\text{s}}\left(\frac{\eta_1}{1-\eta_2}\right)-x_1=
\sum_{n=2}^\infty (v_d \rho)^{n-1}\sum_{n_1=0}^n\frac{n!}{n_1! n_2!}
x_1^{n_1} x_2^{n_2} b_{n_1}\frac{n_1}{n}\sigma_1^{d(n_1-1)}\sigma_2^{dn_2},
\label{33}
\end{equation}
\begin{equation}
\frac{x_2}{1-\eta_1}Z_{\text{s}}\left(\frac{\eta_2}{1-\eta_1}\right)-x_2=
\sum_{n=2}^\infty (v_d \rho)^{n-1}\sum_{n_1=0}^n\frac{n!}{n_1! n_2!}
x_1^{n_1} x_2^{n_2} b_{n_2}\frac{n_2}{n}\sigma_1^{dn_1}\sigma_2^{d(n_2-1)}.
\label{34}
\end{equation}

This allows one to obtain $Z_{\text{m}}(\eta )$ in terms of $Z_{\text{s}}$
evaluated at $\eta $, $\eta _{1}/(1-\eta _{2})$ and $\eta _{2}/(1-\eta _{1})$. 
After some algebra, we arrive at Eqs.\ (\ref{28})--(\ref{28.4}).


\begin{references}

\bibitem[*]{andres}  Permanent address: Departamento de F\'{\i}sica,
Universidad de Extremadura,
E-06071 Badajoz, Spain; Email address: andres@unex.es

\bibitem[\dagger]{santos}  Permanent address: Departamento de F\'{\i}sica,
Universidad de Extremadura,
E-06071 Badajoz, Spain; Email address: santos@unex.es

\bibitem[\ddagger]{mariano}  Also Consultant at Programa de Simulaci\'{o}n
Molecular del Instituto Mexicano del Petr\'{o}leo; Email address:
malopez@servidor.unam.mx

\bibitem{vdW1873}  {\sc van der Waals, J. D., }{\em Ph. D. dissertation},
1873, Leiden. { Translated in {\sc Rowlinson, J. S.}, 1988, {\em Studies in 
Statistical Mechanics}, vol. XIV (Amsterdam: North-Holland).}

\bibitem{KO}  {\sc Kammerlingh Onnes, H.,} 1909, {\em Comm. Leiden}, {\bf 71}.
 
\bibitem{MS69}  See, for instance, {\sc Mason, E. A., and Spurling, T. H.},
1969, {\it The virial equation of state} (New York: Pergamon).

\bibitem{MM40}  {\sc Mayer, J. E., and Mayer, M. G.}, 1940, {\it 
Statistical Mechanics} (New York: Wiley), Chapter 13.

\bibitem{vvc1}  Early  analytical results on the virial coefficients of
hard-sphere fluids ($d=3$) up to the fourth include {\sc J\"{a}gger, G.},
1896, {\em Sitzber. Akad. Wiss. Wien, Math. Natur.-w Kl.} (Pt. 2a), {\bf 105}, 
15, and {\sc Boltzmann, L.}, 1896, {\em Sitzber. Akad. Wiss. Wien, Math.
Natur.-w Kl.} (Pt. 2a), {\bf 105}, 695; 1899, {\em Proc. Kon. Acad. V.
Wetensch., Amsterdam}, {\bf 1}, 398. Those of hard-disc fluids ($d=2$) are
due to {\sc Tonks, L.,} 1936, {\em Phys. Rev.}, {\bf 50}, 955; {\sc Rowlinson, 
J. S.}, 1963, {\em Molec. Phys.}, {\bf 7}, 593, and {\sc Hemmer, P. C.}, 1965, 
{\em J. chem. Phys.}, {\bf 42}, 1116. Numerical calculations of higher 
coefficients up to $n=7$ for both hard discs and hard spheres can be found in 
{\sc Ree, F. H., and Hoover, W. G.}, 1967, {\em J. chem. Phys.}, {\bf 46}, 4181; 
{\sc Kratky,
K. W.}, 1976, {\em Physica} A, {\bf  85}, 607; 1977, {\em Physica} A, {\bf 
87}, 584; 1982, {\em J. statist. Phys.}, {\bf 27}, 533; 1982, {\em J. statist.
Phys.}, {\bf 29}, 129, and {\sc Janse van Rensburg, E. J., and Torrie, G.
M.}, 1992, {\em J. Phys.} {A}, {\bf 26}, 943. The most recent
calculations for these dimensions up to $n=8$ are those of Ref.\ \cite{JvR93}. 
In the case of hard hyperspheres in 4D and 5D, the available results
are those of Refs.\ \cite{LB82} and \cite{J82}.

\bibitem{JvR93}  {\sc Janse van Rensburg, E. J.}, 1993, {\em J. Phys.} A, 
{\bf 26}, 4805.

\bibitem{LB82}  {\sc Luban, M., and Baram, A.}, 1982, {\em J. chem. Phys.}, 
{\bf 76}, 3233.

\bibitem{J82}  {\sc Joslin, C. J.,} 1982, {\em J. chem. Phys.}, {\bf 77},
2701.

\bibitem{KM75}  {\sc Kihara, T., and Miyoshi, K.}, 1975, {\em J. statist.
Phys.}, {\bf 13}, 337.

\bibitem{BXHB88}  {\sc Barrat, J.-L., Xu, H., Hansen, J.-P., and Baus, M.},
1988, {\em J. Phys.} C, {\bf 21}, 3165.

\bibitem{SFG96}  {\sc Saija, F., Fiumara, G., and Giaquinta, P. V.}, 1996, 
{\em Molec. Phys.}, {\bf 87}, 991.

\bibitem{EACG97}  {\sc Enciso, E., Almarza, N. G., Calzas, D. S., and 
Gonz\'{a}lez, M. A.}, 1997, {\em Molec. Phys.}, {\bf 92}, 173.

\bibitem{SFG97}  {\sc Saija, F., Fiumara, G., and Giaquinta, P. V.}, 1997, 
{\em Molec. Phys.}, {\bf 92}, 1089.

\bibitem{EAGB98}  {\sc Enciso, E., Almarza, N. G., Gonz\'{a}lez, M. A., and
Bermejo, F. J.}, 1998, {\em Phys. Rev.} E, {\bf 57}, 4486.

\bibitem{W98a}  {\sc Wheatley, R. J.}, 1998, {\em Molec. Phys.}, {\bf 93},
675.

\bibitem{WSG98}  {\sc Wheatley, R. J., Saija, F., and Giaquinta, P. V.},
1998, {\em Molec. Phys.}, {\bf 94}, 877.

\bibitem{B98}  {\sc Blaak, R.}, 1998, {\em Molec. Phys.}, {\bf 95}, 695.

\bibitem{W99a}  {\sc Wheatley, R. J.}, 1999, {\em Molec. Phys.}, {\bf 96},
1805.

\bibitem{EAGB00}  {\sc Enciso, E., Almarza, N. G., Gonz\'{a}lez, M. A., and
Bermejo, F. J.}, 2001, unpublished. We want to thank these authors for
providing their results to us prior to publication.

\bibitem{EW84}  {\sc Erpenbeck, J. J., and Wood, W. W.}, 1984, {\em J. statist.
Phys.}, {\bf 35}, 321.

\bibitem{EL85}  {\sc Erpenbeck, J. J., and Luban, M.}, 1985, {\em Phys. Rev.}
{A}, {\bf 32}, 2920.

\bibitem{S94}  {\sc Sanchez, I. C.}, 1994, {\em J. chem. Phys.}, {\bf 101},
7003.

\bibitem{CB98}  {\sc Coussaert, T., and Baus, M.}, 1998, {\em J. chem. Phys.}, 
{\bf 109}, 6012.

\bibitem{MV99}  {\sc Malijevsk\'{y}, A., and Veverka, J.}, 1999, {\em Phys.
Chem. Chem. Phys.}, {\bf 1}, 4267.

\bibitem{CS69}  {\sc Carnahan, N. F., and Starling, K. E.}, 1969, {\em J.
chem. Phys.}, {\bf 51}, 635.

\bibitem{EOSDE}  Particular examples for hard discs are discussed in {\sc 
Santos, A., L\'{o}pez de Haro, M., and Yuste, S. B.}, 1995, {\em J. chem.
Phys.}, {\bf 103}, 4622. The most important equations of state for hard spheres 
have been
recently examined in Ref.\ \cite{MV99}.

\bibitem{BC87}  {\sc Baus, M., and Colot, J. L., }1987, {\em Phys. Rev.} A, 
{\bf 36}, 3912.

\bibitem{SM90}  {\sc Song, Y., and Mason, E. A.,} 1990, {\em J. chem. Phys.}, 
{\bf 93}, 686.

\bibitem{LM90}  {\sc Luban, M., and Michels, J. P. J.}, 1990, {\em Phys. Rev.
} A, {\bf 41}, 6796.

\bibitem{BMC99}  {\sc Bishop, M., Masters, A., and Clarke, J. H. R.}, 1999, 
{\em J. chem. Phys.}, {\bf 110}, 11449.

\bibitem{S00}  {\sc Santos, A.,} 2000, {\em J. chem. Phys.}, {\bf 112},
10680.

\bibitem{SYH99}  {\sc Santos, A., Yuste, S. B., and L\'{o}pez de Haro, M.},
1999, {\em Molec. Phys.}, {\bf 96},1.

\bibitem{W98b}  {\sc Wheatley, R. J.}, 1998, {\em Molec. Phys.}, {\bf 93},
965.

\bibitem{W99b}  {\sc Wheatley, R. J.}, 1999, {\em J. chem. Phys.}, {\bf 111}, 
5455.

\bibitem{CCHW00}  {\sc Cao, D.,  Chan, K.-Y., Henderson, D. and Wang, W.},
2000, {\em Molec. Phys.}, {\bf 98}, 619.

\bibitem{AGH}  {\sc Gonz\'{a}lez Melchor, M., Alejandre, J., and L\'{o}pez de Haro,
M.}, 2001, {\em J. chem. Phys.}, {\bf 114}, 4905.

\bibitem{S99}  {\sc Santos, A.}, 1999, {\em Molec. Phys.}, {\bf 96}, 1185.
[Erratum: 2001, {\em Molec. Phys.}, {\bf 99}, 617.]

\bibitem{THM99}  {\sc Tukur, N. M., Hamad, E. Z., and Mansoori, G. A.},
1999, {\em J. chem. Phys.}, {\bf 110}, 3463.

\bibitem{SPT}  {\sc Reiss, H., Frisch, H. L., and Lebowitz, J. L.}, 1959, 
{\em J. chem. Phys.}, {\bf 31}, 369; {\sc Helfand, E., Frisch, H. L., and
Lebowitz, J. L.}, 1961, {\em J. chem. Phys.}, {\bf 34}, 1037.

\bibitem{H95}  {\sc Hamad, E. Z.}, 1995, {\em J. chem. Phys.}, {\bf 103},
3733.

\bibitem{H94}  {\sc Hamad, E.}, 1994, {\em J. chem. Phys.}, {\bf 101}, 10195.

\bibitem{BS00}  { For additional consistency conditions on the contact values of
$g_{ij}$, see {\sc Barrio, C., and Solana, J. R.}, 2000, {\em J. chem. Phys.}, 
{\bf 113}, 10180.}


\bibitem{BS99}  {\sc Barrio, C., and Solana, J. R.}, 1999, {\em Molec. Phys.}, 
{\bf 97}, 797. { For the adaptation of this procedure to the case of
hard-disk mixtures, see
{\sc Barrio, C., and Solana, J. R.}, 2001, {\em  Phys. Rev.} E, 
{\bf 63}, 011201.}


\bibitem{note}  We want to point out that the approximation used in Ref.\ 
\cite{MV99} is somewhat different to the one used here, since apparently the
number of significative figures in the value of the virial coefficients is
not the same in both instances. We have checked that the present
approximation yields practically identical results.

\bibitem{BMCSL}  {\sc Boubl\'{i}k, T.}, 1970, {\em J. chem. Phys.}, {\bf 53}, 
471; {\sc Mansoori, G. A., Carnahan, N. F., Starling, K. E., and Leland,
T. W.}, 1971, {\em J. chem. Phys.}, {\bf 54}, 1523.

\bibitem{BMLR96}
{\sc Baro\v{s}ov\'a, A., Malijevsk\'{y}, A., Lab\'{\i}k, S., and Smith, W. R.}, 
1996, {\em Molec. Phys.}, {\bf 87}, 423.

\end{references}
\end{document}